# Cepstral Analysis to accelerate Green-Kubo thermal conductivity calculations of Metal-Organic Frameworks

*Florian P. Lindner*[a], *Egbert Zojer**[a], *Sandro Wieser**[b]

[a] Institute of Solid State Physics, Graz University of Technology, 8010 Graz, Austria

[b] Institute of Materials Chemistry, TU Wien, Vienna, Austria

**E-mail:** egbert.zojer@tugraz.at, sandro.wieser@tuwien.ac.at

ABSTRACT: Metal-organic frameworks (MOFs) are promising porous materials for applications such as gas storage and separation, where heat transport can critically affect device performance. However, reliable computational prediction of their thermal conductivities remains challenging. In particular, equilibrium molecular-dynamics-based Green-Kubo (GK) simulations, as the most widely used approach, are severely affected by statistical noise. Moreover, they rely on multiple ambiguous, user-defined parameters, which hinder transferability and automation. Here, we demonstrate for metal-organic frameworks that cepstral analysis in combination with GK simulations provides a robust route to massively mitigate these problems, while simultaneously reducing the required sampling times. This is shown for three prototypical frameworks, MOF-5, HKUST-1, and ZIF-8, employing machine-learned moment tensor potentials trained on DFT reference data. In contrast to conventional, direct GK analysis, which shows erratic convergence and strong sensitivity to ad hoc choices of parameters, the cepstral approach yields stable results across a wide range of correlation lengths and achieves convergence within about 1–2 ns of total sampling time. This establishes cepstral analysis base Green-Kubo simulations combined with machine-learned potentials as an efficient, reproducible and automation-ready framework for near-ab initio accuracy prediction of thermal transport in MOFs and other complex low-thermal-conductivity materials.

# 1. Introduction

Metal–organic frameworks (MOFs) have garnered intense research interest due to their high porosity and structural tunability. The latter arises from their hybrid organic-inorganic architecture, which combines organic linker molecules with metal ion nodes or clusters, leading to an effectively unlimited number of possible structures.[1] Numerous innovative applications have been proposed for these materials, ranging from water harvesting[2] and nano-shock absorption[3,4] to iontronic devices[5,6]. Still, their most prominent field of use remains gas storage and separation.[7,8] For such devices, adsorption and diffusion properties are widely studied. In contrast, thermal transport in MOFs remains comparatively underexplored despite its practical importance, especially in adsorption-based storage systems.[9] Here, heat released during gas uptake can induce significant temperature spikes (ranging from 100K to 250K) that reduce the device efficiency, whereas rapid desorption may cause cooling that hinders complete release.[9–11] Reliable knowledge of thermal conductivity is, therefore, essential for assessing device-level performance and for guiding materials design.

Experimental measurements of the intrinsic thermal conductivity of MOFs remain scarce, primarily due to the difficulty of preparing large, defect-free single crystals. To the best of our knowledge, single-crystal thermal conductivity measurements have only been reported for a few prototypical systems, namely MOF-5,[12] HKUST-1,[13] and ZIF-8.[14] Given the vast number of synthesized and hypothetical MOF structures and the experimental difficulties in accurately characterizing their heat transport properties, reliable computational tools appear as an indispensable tool for systematically exploring thermal transport.

Because most MOFs are electrical insulators, classical molecular dynamics (MD) simulations are particularly well suited for studying thermal transport in these systems. MD-based approaches for simulating thermal conductivities include non-equilibrium simulations employing, for example, the "non-equilibrium molecular dynamics" technique (NEMD).[15,16] As it relies on determining the heat flux for a given temperature gradient (or vice versa), it corresponds to an atomistic analogue of Fourier's law.[16] The disadvantages of NEMD simulations are that they require the introduction of artificially large temperature gradients and that NEMD simulations suffer from severe finite-size effects, which are commonly treated by more or less heuristic finite-size extrapolation schemes.[17,18] This can lead to systematic errors that cannot easily be quantified and, the finite-size extrapolation requires the time-consuming simulations of multiple, often huge supercells.[15,18–20] This makes NEMD simulations, for example, less suited for high-throughput studies.

A conceptually different approach is equilibrium molecular dynamics (EMD) within the Green-Kubo (GK) formalism.[21–24] It relates the thermal conductivity of a material to equilibrium fluctuations of the microscopic heat flux via linear response theory.[23,24] The GK approach is formally exact and avoids explicit perturbations. Moreover, compared to NEMD simulations, finite-size effects are generally less severe.[18] Although this view is widely accepted, noticeable size dependencies have been reported for MOF systems by Ying et al.[25] A reason for that is that the GK method operates under fully periodic boundary conditions and does not require the explicit creation of hot and cold reservoirs. This avoids introducing artificial thermostat boundaries, at which phonon scattering occurs.[18]

Still, despite being conceptually exact, obtaining converged thermal conductivities from direct GK evaluations can be very challenging in practice. The root cause for this is that the exact relations from linear response theory have to be approximated by a finite, discrete time series of the heat flux. This causes significant statistical noise, which contaminates all appearing quantities. This applies in particular to the heat flux auto correlation function (HFACF), which has to be calculated to obtain thermal conductivities (see below). The challenge for materials with high thermal conductivities is that the HFACF decays only slowly. Thus, it must be sampled over long timescales to capture the full correlation.[26] At first sight, this makes the GK method especially interesting for MOFs, as here the HFACF is expected to decay very quickly due to the materials' low thermal conductivities.[27] However, in such cases, the physically relevant part of the HFACF typically decays within only a few picoseconds, after which the signal becomes dominated by statistical noise originating from the finite trajectory length. This leads to a different kind of convergence difficulty, as will be shown below.

Due to this "noise challenge", conventional GK workflows rely on user-defined parameters such as smoothing windows, correlation lengths, and extraction criteria, which are often chosen system-specifically and whose choice can introduce significant uncertainties and biases.[28–30] This makes it challenging to use the GK approach for the reliable and automated simulation of thermal conductivities.[27]

This problem is overcome in the present study by augmenting the GK framework for evaluating thermal conductivities of MOFs with cepstral analysis.[31] Cepstral analysis is a signal processing technique in which the logarithm of a spectrum is transformed back into a time-like domain, the so-called cepstrum. In simple terms, this makes it possible to separate slowly varying spectral features from rapidly fluctuating noise-like contributions. Originally, cepstral methods were developed for applications such as echo detection and signal deconvolution.[31] In the context of GK-based heat-transport simulations, cepstral analysis substantially reduces the required trajectory lengths to obtain converged thermal conductivity coefficients, while providing a statistically rigorous estimate of their uncertainties. Moreover, it removes the strong dependence on user-defined, ad hoc parameters that often limits the reproducibility of conventional, direct GK evaluations. In passing, we note that the present paper is not the first time cepstral analysis has been applied in the context of GK simulations. The application of cepstral analysis to GK transport calculations was pioneered by Baroni and co-workers,[32–35] albeit for materials fundamentally different from the MOFs considered here, namely elemental and molecular fluids (liquid Ar, $H_2O$) and glassy as well as simple crystalline solids (a-$SiO_2$, MgO). Also, cepstral analysis was recently applied to InAs nanowires, where the application of the approach for some systems turned out to be less straightforward due to their significantly higher thermal conductivity.[26]

While we will show that exploiting cepstral analysis massively reduces the required sampling times for GK simulations, accurate force descriptions remain essential for reliable thermal transport predictions. The recent emergence of machine-learned interatomic potentials (MLIPs) makes it possible to perform nanosecond-scale MD simulations with near ab initio accuracy at a tiny fraction of the computational cost of DFT. In the present work, we employ the moment tensor potential (MTP) model,[36,37] which offers an excellent compromise between computational efficiency and accuracy (although at the expense of transferability). By combining MLIP-based GK simulations with cepstral analysis, we will illustrate how robust thermal conductivity estimates at a substantially reduced computational expense can be achieved for prototypical MOF systems

## 1.1. The studied systems

We focus on three prototypical systems: MOF-5,[38] HKUST-1,[39] and ZIF-8,[40] with their structures shown in Figure 1. This choice is motivated by the availability of single-crystal experimental data[12–14] and extensive prior theoretical investigations of their thermal transport properties.[25,41–47] MOF-5 and HKUST-1 are carboxylate-based frameworks that at room temperature crystallize in the cubic space group Fm-3m (225), with lattice constants of 25.8 Å and 26.3 Å, respectively.[39,48] MOF-5 consists of $Zn_4O$ secondary building units connected by 1,4-benzenedicarboxylate (BDC) linkers and contains 424 atoms in its conventional unit cell. HKUST-1 is based on Cu paddlewheel nodes linked by benzene-1,3,5-tricarboxylate (BTC) units and comprises 624 atoms per conventional unit cell. ZIF-8, a representative member of the zeolitic imidazolate framework family, crystallizes in the cubic space group I-43m (217) with a lattice constant of 17.0 Å.[49] It is composed of tetrahedral $ZnN_4$ nodes connected by 2-methylimidazolate linkers and contains 276 atoms in the conventional unit cell. While MOF-5 and HKUST-1 are widely studied for gas storage and catalytic applications,[50–53] ZIF-8 has attracted significant attention in areas such biomolecular encapsulation.[54–56]

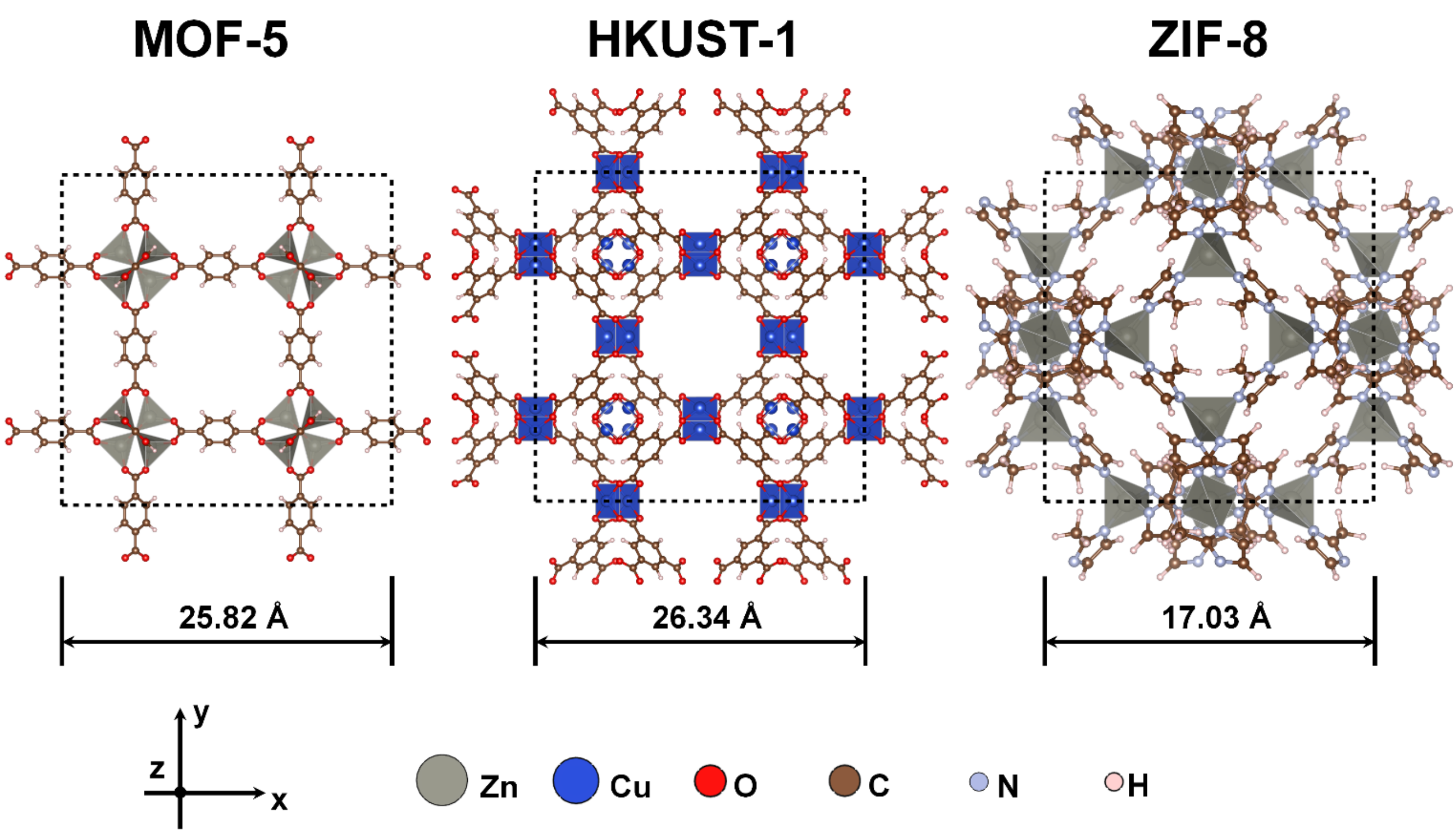

Figure 1. Structures of the investigated MOFs together with their experimental conventional lattice constants.[48,39,49] The dashed black lines indicate the conventional crystallographic unit cells of the respective frameworks, comprising 424 atoms for MOF-5, 624 atoms for HKUST-1 and 276 atoms for ZIF-8.

# 2. Computational Details

## 2.1. DFT calculations

All DFT calculations (used for generating reference data for training the machine-learned potentials) were performed using the VASP code.[57–59] Unless otherwise specified, a plane-wave energy cut-off of 900 eV and a Γ-point-only k-point sampling were employed. The latter is justified by the flat electronic bands of the studied MOFs.[60,61] Exchange-correlation effects were treated using the generalized gradient approximation (GGA) with the Perdew-Burke-Ernzerhof (PBE) parametrization.[62,63] To account for dispersion interactions, we applied Grimme's D3 correction scheme[64] with Becke-Johnson damping.[65] The parametrization of the machine-learned interatomic potentials (MLPs) used in this study follows a combined active and passive learning framework introduced and extensively validated previously.[60] For completeness, we briefly summarize the key steps of this workflow and explain the basic settings for passive learning of the used Moment Tensor Potentials (MTPs).[36,37]

## 2.2. VASP active learning – Generation of DFT training data

The active learning strategy used to generate DFT training data for the VASP machine-learned potentials (VMLPs)[66,67] follows the suggested standard paradigm of on-the-fly data acquisition during MD simulations.[66] In our case, active learning was conducted during MD simulations, starting from an initial temperature of 50 K and linearly increasing the temperature to 900 K over the course of 50,000 MD steps, with a timestep of 0.5 fs. All other machine learning settings were kept at their default values (VASP version 6.4.1). The procedure begins with training an initial kernel-based VMLP on a small set of structures (~ 10) sampled from short ab initio MD simulations. This preliminary model is then used to propagate the MD trajectory. At each step, a Bayesian error estimate is used to assess the reliability of the predicted atomic forces. If the predicted uncertainty exceeds a dynamically adapted threshold, the corresponding structure is recomputed using DFT. The parameters that control the adaption of the Bayesian error threshold were kept at their default values. Finally, the resulting structure is added to the reference dataset, and the VMLP is retrained on the expanded set. As the set of reference structures grows, the model becomes more accurate, and DFT calls are triggered less frequently. Using this protocol on each MOF individually, we obtained 974, 1431 and 1809 DFT-evaluated reference structures for MOF-5, HKUST-1, and ZIF-8, respectively, each providing DFT calculated energies, forces, and stresses. These reference data are then used in a passive learning scheme to obtain MTP models.

## 2.3. Passive learning of Moment Tensor Potentials

Using the DFT reference structures from active learning described in the previous section, we trained a series of MTPs with the MLIP-2 package.[36] MTPs represent the total energy of a system as a linear combination of moment tensor descriptors,[37] fixed basis functions that capture the local atomic environment. The complexity of the potential is governed by a level parameter, which we set to 18; the radial basis set size was selected to be 12. To enhance model accuracy without compromising computational efficiency in the MD runs, we assign separate parameter sets to atoms in specific chemical environments (for a complete list of atom types see **Supporting Information**). As previously demonstrated, this strategy improves prediction quality while maintaining the inherent speed advantages of MTPs.  The model parameters were optimized by minimizing a weighted cost function that incorporates DFT-calculated energies, forces, and stress tensors. We assigned dimensionless weights of 1.0, 0.01, and 1.0 to the energy, force, and stress components, respectively. Given the stochastic nature of the training process, we trained 10 independent MTP models. From these, we selected the one that exhibited the best balance between accuracy and thermal stability during test MD simulations, as explained in detail in the **supporting information**.

## 2.4. Assessment of the used machine-learned potentials

The predictive quality of the MTP models employed here has been extensively validated in our previous work for a range of MOF systems and properties.[60] As the present study follows identical training protocols and simulation conditions, we focus on a targeted validation relevant to the current simulations. The DFT reference dataset used for validation was generated by sampling structures along VMLP-based MD trajectories in an NpT ensemble at 100 K, 300 K, and 500 K. From each trajectory, configurations were extracted every 10,000 MD steps, resulting in 150 diverse structures across the full temperature range. Single-point DFT calculations were then performed for each structure to obtain energies, forces, and stresses. Thus, in total, the used validation set consists of 450 DFT calculated structures. Figure 2a shows the force comparison plots for the three MTPs used to model the MOF systems, demonstrating excellent agreement with the DFT reference data. For the 300 K validation structures, root mean square errors (RMSEs) for the forces are 47 meV $Å^{-1}$ for MOF-5, 41 meV $Å^{-1}$ for HKUST-1, and 66 meV $Å^{-1}$ for ZIF-8. As expected, the RMSEs increase with the absolute forces at higher temperatures. RMSE values at 100 K and 500 K, as well as for the combined validation set, are provided in the **Supporting Information**. The corresponding histogram of force deviations, depicted in figure 2b, further confirms the narrow force-error distributions.

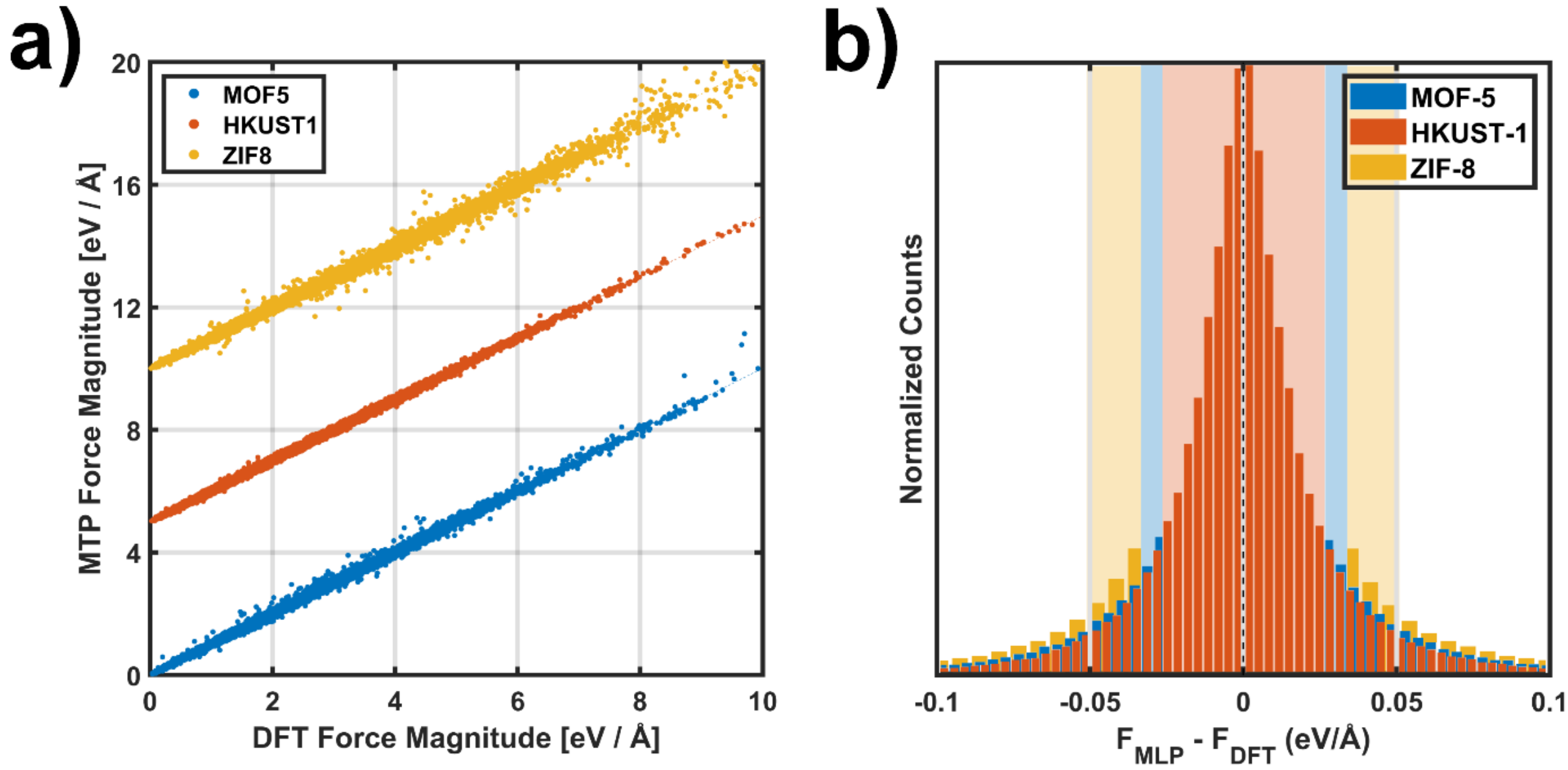

Figure 2. Validation of the MTPs used to model the three MOF systems against a test set comprised of structures sampled from NpT-MD trajectories at 100K, 300K and 500K.(a) Parity plot comparing DFT reference forces with forces predicted by the MTPs. The forces were offset by in 5 $eV\,Å^{-1}$ intervals for the different systems. (b) Histogram of force deviations (ΔF = $F_{MTP}$ - $F_{DFT}$), which displays the narrow error distribution of the used models; shaded regions correspond to one standard deviation.

Among the three systems, the MTP for ZIF-8 exhibits the broadest distribution. We attribute the slightly larger errors for ZIF-8 to the presence of methyl groups, whose rotational degrees of freedom are associated with very shallow energy barriers, making them more challenging to describe accurately.

## 2.5. Computational details for Green-Kubo simulations and Cepstral analysis

All molecular dynamics (MD) simulations were carried out using LAMMPS.[68] Especially in the case of complex many-body potentials, like MLIPs, care must be taken on how the virial part of the Hardy heat flux (5) is evaluated. While for pair-potentials one has:[69]

$$\frac{\partial U_i^{pair}}{\partial \mathbf{r_{ji}}} = - \frac{\partial U_j^{\mathrm{pair}}}{\partial \mathbf{r_{ij}}} \tag{1}$$

Equation (1) is in general not true for MLIPS, due to their many-body character. As was recently discussed,[69] the implicit assumption of the asymmetry condition (1) in the calculation of the Hardy flux for MLIPs leads to inconsistencies, which cause errors in the computed thermal conductivities.[69–72] For the MTPs used throughout this work, this has very recently been fixed by Tai et al.[70] through a modified LAMMPS-MLIP interface.[36,70] The latter was also used for all simulations contained in this work.

In all MD runs, the time step was set to 0.5 fs. The room temperature unit cells used in the GK simulations were obtained in NPT simulations running at 300 K for 175 ps, and employing 2x2x2 supercells of the conventional unit cells shown in Figure 1. The first 10,000 steps were discarded to avoid bias from poorly equilibrated structures. Green-Kubo production runs were performed in the NVE ensemble, following a 30 ps equilibration period in the NVT ensemble. Each trajectory covered 1 ns (corresponding to 2,000,000 steps). Tests varying the supercell-size showed that 3x3x3 supercells of the conventional unit cells yield converged results for each of the three MOF systems (MOF-5, HKUST-1, and ZIF-8). More details on finite size effects can be found in the **Supporting Information**. To improve statistical sampling, up to 10 independent trajectories were generated by reinitializing atomic velocities with different random seeds. For the cepstral analysis, we employed the GK_analysis code package recently published by a co-author of the present study.[26]

# 3. Fundamental Methodological Aspects

Based on the fluctuation dissipation theorem,[73] the GK equations connect spontaneous fluctuations of the heat flux in thermal equilibrium to transport under non-equilibrium conditions. In this chapter, we first summarize the main concepts of the GK theory and then recapitulate how the exact equations are approximated in finite-length MD simulations. This is done to provide a firm basis for a better understanding of the challenges discussed later, which arise from the user-based choices of relevant simulation parameters. To exemplify the situation, conventional GK-based simulations of heat transport in MOF-5 are performed. They also clarify the common workflow of a GK-based heat-transport simulation, illustrating possible pitfalls of the direct GK approach.

## 3.1. Green-Kubo formalism for calculating thermal conductivities

In the GK formalism, a direct link between the fluctuations of the microscopic heat flux vector **J** and the macroscopic thermal conductivity tensor **κ** is established via the following time integral:

$$\boldsymbol{\kappa} = \frac{\mathrm{V}}{\mathrm{k_B T^2}} \int_0^{\infty} \langle \mathbf{J}(\mathrm{t}) \otimes \mathbf{J}(0) \rangle \, \mathrm{dt} \tag{2}$$

Here, V and T are the system volume and temperature and $k_B$ is the Boltzmann constant. The operator $\otimes$ denotes the outer product in $\mathbb{R}^3$. The instantaneous heat flux **J**(t) characterizes the direction and magnitude of energy flow through the system.

In equilibrium, the term $\langle \mathbf{J}(t) \otimes \mathbf{J}(0) \rangle$ corresponds to the heat flux autocorrelation function (HFACF), **C**(t), which represents an ensemble average over equilibrium phase-space configurations weighted by the equilibrium distribution $\rho(\Gamma_0)$:

$$\mathbf{C}(t) = \langle \mathbf{J}(t) \otimes \mathbf{J}(0) \rangle_{eq} = \int \mathbf{J}(\Gamma_t) \otimes \mathbf{J}(\Gamma_0)\, \rho(\Gamma_0)\, d\Gamma_0$$

*( 3 )*

In eq. (3), $\Gamma = (\mathbf{r}_1,\ldots,\mathbf{r}_N,\mathbf{p}_1,\ldots,\mathbf{p}_N)$ denotes a point in phase space and $\rho(\Gamma_t)$ is the equilibrium probability density for the system to occupy the phase-space point $\Gamma$ at time t. The HFACF measures how strongly in equilibrium the flux at time t (with the system in state $\Gamma_t$) remains correlated with a spontaneous flux fluctuation at an arbitrary time $t_0$ (with the the system in state $\Gamma_0$). In view of eq. (2) rapidly decaying correlations, i.e., small values of $\mathbf{C}(t)$, relate to low thermal conductivities and vice versa.

In this work, we focus on cubic systems, in which the thermal conductivity is isotropic and, therefore, consider the averaged HFACF, defined as:

$$C(t) = \frac{1}{3} \langle \mathbf{J}(t) \cdot \mathbf{J}(0) \rangle_{eq}$$

*( 4 )*

Here, the outer product between the heat flux vectors has been replaced by a scalar product. The heat flux $\mathbf{J}(t) \equiv \mathbf{J}(\Gamma_t)$, i.e., the microscopic heat flux evaluated along an MD trajectory, can be decomposed into a convective contribution $\mathbf{J}_{conv.}$, accounting for the convection of energy by particle motion, i.e., mass transport, and a virial contribution $\mathbf{J}_{virial.}$ arising from energy exchange mediated by interatomic forces. Based on the microscopic heat-flux formalism originally introduced by Hardy et al.,[74] for a system of multiple atoms it can be written as:[71]

$$\mathbf{J}(t) = \mathbf{J}_{conv.} + \mathbf{J}_{virial} = \frac{1}{V} \sum_i E_i \mathbf{v_i}(t) + \frac{1}{V} \sum_{j \neq i} \mathbf{r_{ij}} \left( \frac{\partial U_j}{\partial \mathbf{r_{ji}}} \cdot \mathbf{v_i}(t) \right)$$

*( 5 )*

Here, $E_i = 1/2\ m_i\ \mathbf{v}_i^2 + U_i$ denotes the total energy of atom "i", $\mathbf{v}_i$ is the velocity of atom "i" and $U_i$ is its potential energy. The position vector $\mathbf{r}_{ij}=\mathbf{r}_i - \mathbf{r}_j$ is the distance vector between atoms "j" and "i".

Finally, the angle brackets ⟨...⟩ in eq. (2) denote the equilibrium ensemble average. In MD simulations, the flux $\mathbf{J}(t)$ as given by eq. (5) is sampled along an equilibrium trajectory of finite time, $T_{tot}$, yielding a discrete and finite sequence $\mathbf{J}(t_n)$. Here, $t_n = n\Delta t$, with an integer n and with $\Delta t$ representing the time step used in the MD simulation. Under the assumption of ergodicity of the system, the ensemble average in eq. (2) and eq. (3) can be replaced by time averages over different initial times, $t_0$, along a sufficiently long trajectory. For a trajectory consisting of $N_{tot}$ time steps (with a total simulation time of $T_{tot} = N_{tot}\ \Delta t$), the HFACF from eq. (4) can be estimated for a specific time t=tm. This is done via a time average over multiple time origins, $t_i$:

$$C(t_m) = \frac{1}{3(N_{tot} - m)} \sum_{i=0}^{N_{tot}-m-1} \mathbf{J}(t_{i+m}) \cdot \mathbf{J}(t_i) \tag{6}$$

Notably, the larger the value of $t_m$ becomes, the smaller the number of initial times $t_i$, over which one can average before $t_{i+m}$ exceeds the total simulation time $T_{tot}$. In practice, owing to the finite length of the simulation trajectory, the HFACF estimator eq. (6) is inherently noisy, and the situation becomes worse for large $t_m$. A complication in materials with very low thermal conductivities such as MOFs is that $C(t_m)$ typically decays to the level of statistical noise within only a few picoseconds. To mitigate the noise, the raw HFACF obtained from the direct evaluation of eq. (6) is commonly post-processed using a running average, whose window width constitutes a user-defined ad-hoc parameter.

In addition, it is customary to introduce a finite analysis time window $t_{seg}$, beyond which the HFACF is no longer evaluated.[41,75,76] If this is done, in the above considerations, $T_{tot}$ is replaced by $t_{seg}$. Notice that $t_{seg}$, denotes the finite analysis window used for evaluating the HFACF and should not be confused with its decay time. A trajectory of total length $T_{tot}$ is then divided into $M = \lfloor T_{tot} / t_{seg} \rfloor$ segments, which are treated independently. For these segments, individual HFACFs, C(t), are calculated. These are then averaged, which improves statistical sampling. The choice of $t_{seg}$ is made depending on the studied system. In combination with the other choices made along with the workflow introduced below, it influences the resulting thermal conductivity estimates.

## 3.2. Practical challenges: the instructive case of MOF-5

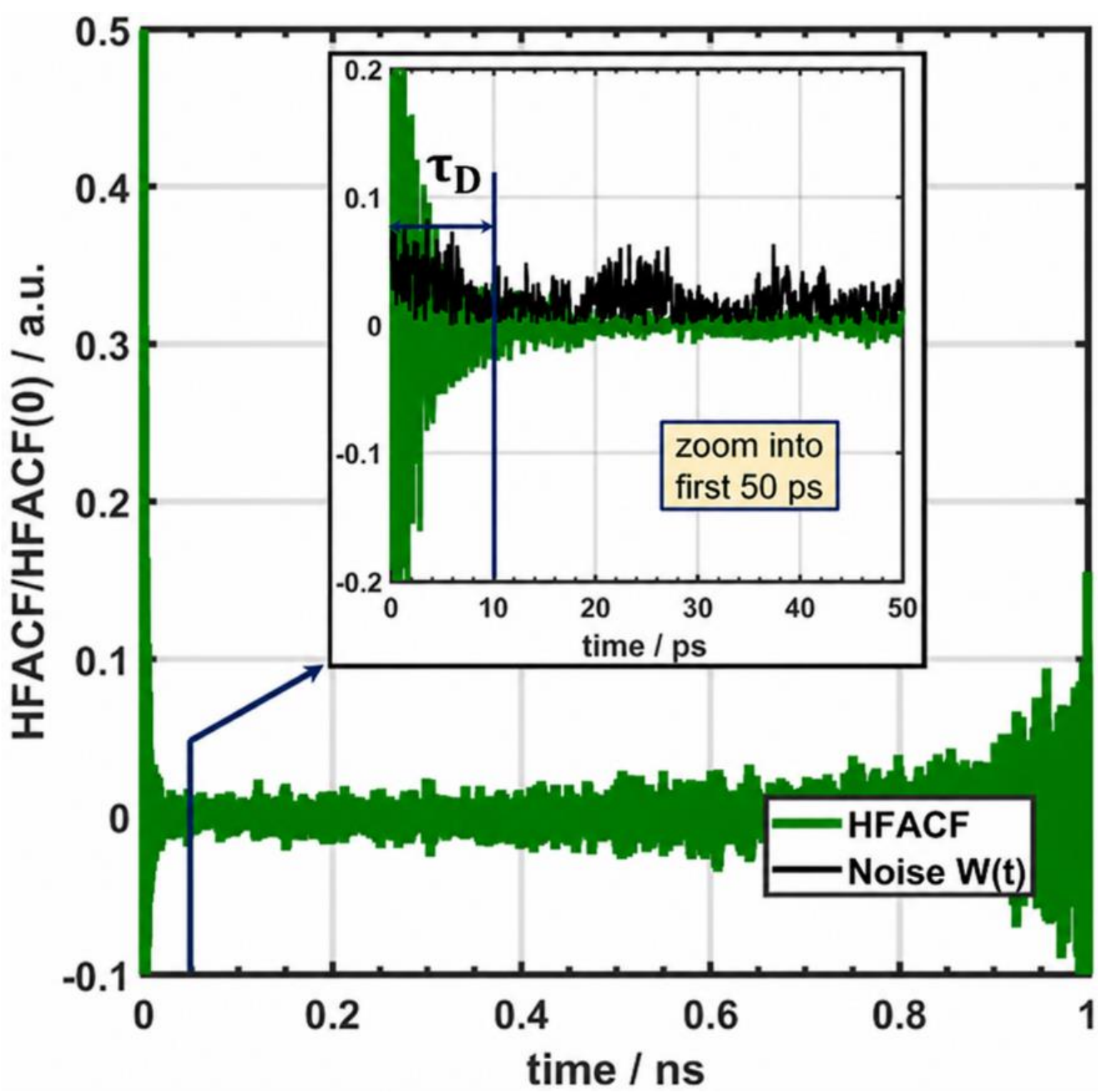

*Figure 3. Example for the time averaged HFACF of MOF-5 calculated using eq. (5) for a full 1 ns trajectory in MOF-5. I.e., here the 1 ns trajectory was not cut into separate analysis windows such that for this example $t_{seg}=T_{tot}$. After an initial decay phase of about 10 ps, the*

*HFACF turns into statistical noise. The noise floor in the zoom plotted as an inset is estimated by the root-mean-square of the flux cross-correlations as defined in (6). For the running average smoothing, a window with a width of 0.5 ps was chosen.*

An example of the shape of a HFACF calculated using equation (6) is shown in figure 3 for the prototypical MOF system MOF-5. The depicted HFACF was obtained for a $T_{tot}=t_{seg}=1$ ns time series of the heat flux. The HFACF is normalized relative to its value at t=0 and a running average with a window width of 0.5 ps is applied to the raw data. Unless stated otherwise, this implicit smoothing of the raw HFACF is applied throughout the remainder of this work. As shown in the inset of figure 3, the HFACF exhibits a rapid initial decay within the first few picoseconds. After this system-specific decay regime, which for MOF-5 extends to approximately 10 ps, the HFACF fluctuates around zero, indicating that the signal has decayed to the level of statistical noise. The magnitude of these fluctuations increases toward the end of the analysis window. This follows directly from equation (6) because for increasing times $t_m$, progressively fewer time origins are available for the averaging. To provide a rough estimate for the magnitude of statistical noise, also the root-mean-square of the heat flux cross-correlation functions, W, is plotted. W is defined as:[76]

$$W(t) = \sqrt{\frac{1}{6}\sum_{\alpha\neq\beta}\langle J_\alpha(t)J_\beta(0)\rangle^2} \quad (7),$$

where the sum runs over all off-diagonal heat flux cross-correlations. In cubic systems at equilibrium, cross-correlations between orthogonal components of the heat flux vanish by symmetry, i.e., $\langle J_\alpha(t)J_\beta(0)\rangle = 0$ for any $\alpha\neq\beta$. Consequently, any finite value of such cross-correlations that has been obtained from an MD trajectory arises solely from statistical fluctuations. We therefore use the quantity W(t) as a very rough estimate of the magnitude of statistical noise that would be expected in the HFACF.[76] In fact, as the root mean square type definition of W(t) favors large values, the noise in W(t) essentially serves as an upper limit to the noise in C(t).

After calculating the HFACF, $C(t_m)$, for a discrete and finite MD trajectory using eq. (6), one can calculate an estimate of the thermal conductivity $\kappa(t_M)$ by numerically integrating $C(t_m)$ up to a certain time, $t_M$. Using the trapezoidal rule, one obtains:

$$\kappa(t_M) = \frac{V}{k_B T^2} \sum_{m=1}^{M} [C(t_m) + C(t_{m-1})] \frac{\Delta t}{2} \tag{8}$$

As eq. (8) represents a running (numerical) integral over the HFACF, statistical noise in $C(t_m)$ is accumulated as the summation proceeds. Thus, as shown in de Sousa Oliveira et al.[28], the integration in eq. (8) exhibits a random-walk like behavior with a variance that grows over time rather than $\kappa(t_M)$ converging to a well-defined value. To suppress heavy fluctuations, it is common practice to apply an additional running average to the integrated κ curve. The width of the used smoothing window constitutes yet another user-defined parameter. The averaged value of $\kappa(t_M)$ for a smoothing window of 0.2 ps is shown as a red line in Figure 4a and in the zoom in Figure 4b. Still, even after smoothing, the extraction of a reasonable thermal conductivity value remains highly non-trivial: if one picked $\kappa(t_M)$ at a too small value of $t_M$, the integral would not yet be converged, while for a too large value of $t_M$ the noise-related problems mentioned above dominate and one might even get negative values of κ. As a consequence, the user again has to make a critical choice, namely the position of the "extraction point", i.e., the value of $t_M$ at which the "final" thermal conductivity is determined.

To guide the decision for the choice of the extraction point, several ad-hoc criteria have been described in literature: one of these is the so-called first-dip criterion.[77] The first-dip criterion can, in principle, provide reasonable estimates when combined with appropriate denoising procedures of the HFACF, like gauge fixing.[22] However, as the latter mostly removes high-frequency noise components,[26] the first dip criterion still might be of limited practical value in low-conductivity MOF systems.

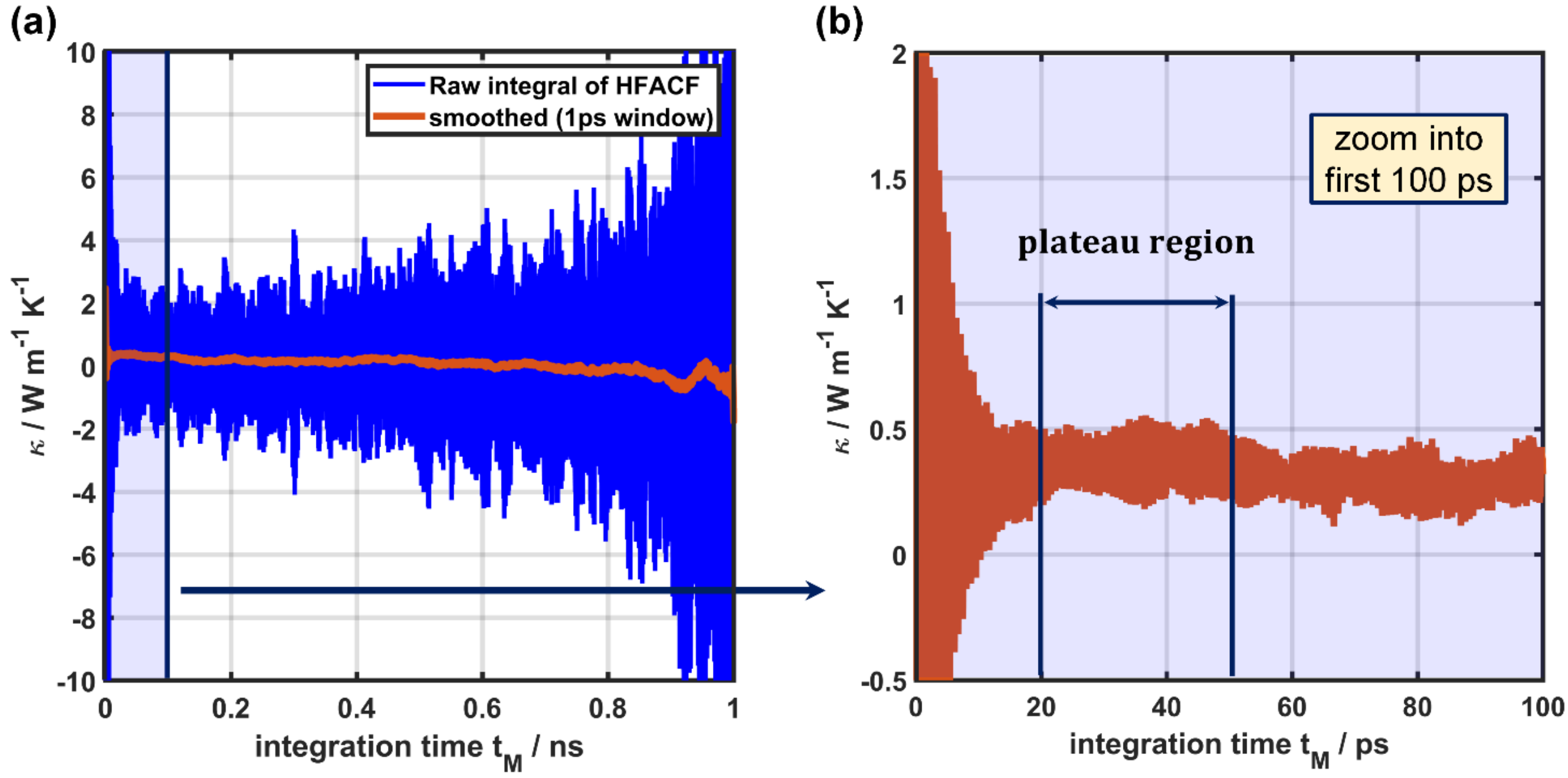

*Figure 4. Numerical integration of the HFACF obtained from the 1 ns trajectory for MOF-5, depicted in Figure 3. (a) Integral over the time averaged HFACF for up to 1 ns without (blue curve) and with 0.2 ps and running average smoothing (red curve). (b) Zoom into the first 100 ps for the smoothed $\kappa$ integral. In this specific simulation, a relatively well-resolved plateau region is observed. It is noted that the width of the smoothing window as well as the identification of a plateau region are choices made by the user.*

An alternative approach, applied, for example in Refs.[27,41] is identifying a plateau region. Nothing like a plateau is visible for the raw integral of $\kappa(t_M)$, as seen in Figure 4a. However, after smoothing, the zoom shown in Figure 4b appears to reveal a plateau in the region between approximately 20 - 50 ps. As mentioned in the Computational details, ten independent runs of 1 ns were performed. Notably, such a plateau is not observed in all cases, due to the random walk-like behavior of the cumulative integral eq. (8). This is illustrated in Figure S4 of the **Supporting Information.**

The above considerations show that for a typical workflow of a GK thermal conductivity calculation, multiple ad-hoc choices have to be made by the user, including the size of the running average windows, the choice of the analysis window length $t_{seg}$, and the identification of suitable plateau regions to extract actual $\kappa$ values. Besides careful human inspection, obtaining statistically reliable and reproducible results, often requires very long total simulation times, in some cases up to tens of nanoseconds, as only then it is possible to average over a sufficiently large number HFACFs.[75,76] Thus, for example, in their recent paper on CALF-20 Mandal et al.[75] averaged HFACFs over at least 16 independent runs requiring ~40 ns of simulation time. Such a requirement is particularly demanding when using anything other than the simplest (and highly unreliable) force fields and when studying MOFs, which typically converge for large supercells containing several hundred atoms. It is, therefore, not surprising that numerous approaches have been proposed to tackle the noise problem in GK simulations. A common procedure in standard GK literature is to omit the convective part $\mathbf{J}_{conv}$ in the heat flux in equation (5). This approach is justifiable for perfectly crystalline materials with atoms only vibrating around their equilibrium positions. It is, however, not ideally suited as a general strategy for MOFs, as many of their primary technical applications involve guest molecules within their pores, whose convection will influence heat transport.[13,78–81] Moreover, this strategy reduces mostly the high-frequency noise,[26] while for determining the thermal conductivity the low-frequency noise is particularly problematic (see section 3.3). To maintain flexibility, we will retain the full heat flux in all subsequent analyses, even though the actually studied examples do not include guest molecules.

There is also a more general theoretical framework, often referred to as gauge fixing. It exploits the gauge invariance of the microscopic heat fluxes to reduce the noise level in the HFACF.[29]

This is achieved by removing the so-called non-diffusive flux components (i.e., terms whose GK integral vanishes) from the total microscopic heat current.[29,22,75] Unfortunately, also gauge fixing, which has been used, for example by Mandal et al.[75], primarily tackles the high-frequency noise.[26] Finally, there are also strategies, which model the cumulative GK integral as a random-walk, or which apply uncertainty-based statistical analyses to quantify the variance in thermal conductivity.[28,30] In the present contribution, we apply a different approach, which will be described in the next section.

## 3.3. Analyzing Green-Kubo integrals via Cepstral analysis

As was outlined in the last subsection, in principle the GK relation (2) is exact, but a direct application to estimate thermal conductivities can be challenging. As discussed above, the reason for that is that equation (2) can in practice only be calculated from finite time series derived from MD runs, which usually introduces significant statistical noise. Moreover, ad-hoc, often system specific parameters need to be chosen and obtaining a statistically sound estimate of the uncertainty of the predicted thermal conductivity values is difficult. These problems can be largely overcome using cepstral analysis. This method has originally been used to detect echoes or reflections in periodic signals,[31] and more recently has also been used in heat-transport simulations.[26,32–35] Here, we will outline the basic working principle of this approach, which is needed to understand the subsequently presented results. Further mathematical details can, for example, be found in the seminal work of Ercole et al.,[32] who, to the best of our knowledge, were the first to apply this technique to the investigation of GK transport integrals. The central object for the cepstral analysis of GK simulations is the power spectrum $S(\omega)$ of the heat flux. Applying the Wiener-Khinchin theorem,[82,83] it can be written as the Fourier transform of the HFACF:

$$S(\omega) = \int_{-\infty}^{\infty} C(t)\, e^{-i\omega t}\, dt$$

(9)

From (9) it follows directly that the thermal conductivity cannot only determined from the GK integral over the HFACF in the time domain, but it is also given by the zero-frequency limit of the power spectrum:[26,32]

$$\kappa = \frac{V}{2k_B T^2} S(\omega \to 0)$$

(10)

In every practical case, the HFACF appearing in (9) has to be estimated again from a discrete and finite time series of the heat flux $J(t_n)$ sampled during an equilibrium MD simulation. Thus, it is necessary to calculate the discrete Fourier transform of the MD sampled HFACF as given in (6). As a consequence, the resulting power spectrum $S(\omega_k)$ is defined at the discrete frequencies $\omega_k = (2\pi\, k)/(N_{tot}\, \Delta t)$. Due to this, also the noise contained in the HFACF $C(t_m)$ will be carried over into the frequency domain. In Figure 5, this is illustrated by the corresponding power spectrum of MOF-5 calculated from the HFACF from Figure 3.

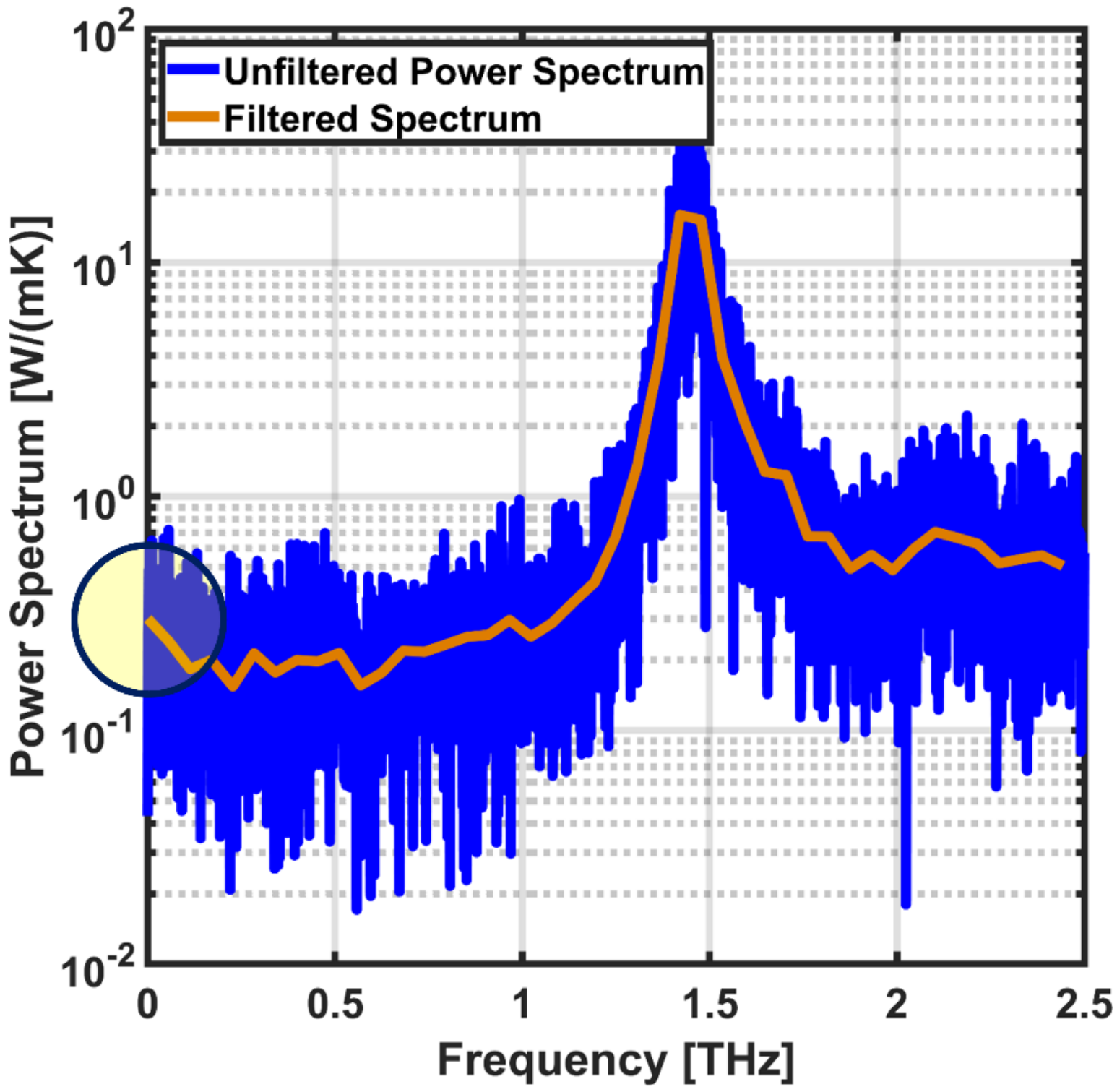

*Figure 5. Power spectrum of the HFACF for MOF-5 as provided in Figure 3. The blue line shows the raw spectrum obtained from the HFACF. The denoised spectrum using cepstral analysis is depicted in orange. The yellow shaded circle is centered at the zero-frequency limit.*

The power spectrum in Figure 5 shows the severe noise problem caused by the finite time sampling of the heat flux persists. In the crucial low-frequency region (highlighted by a yellow circle), the values of S(ω) fluctuate over approximately an order of magnitude, making a direct evaluation of equation (10) to obtain $\kappa$ futile. Notably, the denoising strategies mentioned at the end of the previous section do not reduce the noise level of the spectral function in the (most relevant) low-frequency regime, as shown for the example of InAs nanowires in Wieser et al.[26]. Thus, to deal with the low frequency noise in the power spectrum, Ercole et al.[32] proposed to use cepstral analysis. This requires the calculation of the so-called "Cepstrum", comprising the cepstral coefficients $C_n$. They are the sequence of coefficients of the inverse (discrete) Fourier transformed logarithm of the power spectrum $S(\omega_k)$. The crucial observation by Ercole et al.[32]

has been that cepstral coefficients $C_n$ for large n correspond to noise in the power spectrum. Hence, when keeping only an optimal number P* of cepstral coefficients $C_n$ and setting:

$$C_n = 0, \ \forall \ |n| > P^*$$

(11)

In effect, this removes rapidly varying features from the logarithm of the power spectrum. The reconstructed spectrum is therefore smoothed and less affected by statistical noise, particularly in the low-frequency region that determines the thermal conductivity, see Figure 5. Importantly, the optimal number P* of cepstral coefficients can be obtained from a deterministic criterion, namely by minimizing the second order Akaike information criterion, $AIC_c$,[84] which is a function of the number of cepstral coefficients P:[26,32]

$$P^* = \arg\min \{AIC_c(P)\}$$

(12)

The exact mathematical form of AICc(P) can be found in the Supporting Information. In other words, P* corresponds to the number of cepstral coefficients for which the Akaike information criterion attains its minimum. As a last step, the thermal conductivity can be obtained from the zero-frequency value of the Power spectrum derived from the truncated Cepstrum. Interestingly, the cepstral method also provides access to the statistical uncertainty $\Delta\kappa_{Cep}$ of the obtained $\kappa_{Cep}$ values.[26,32]

$$\Delta\kappa_{Cep} \propto \frac{4P^* - 2}{N_{tot}}$$

(13)

Relation (13) shows that $\Delta\kappa_{Cep}$ increases with the number of cepstral coefficients, P*, and decreases with the trajectory length, $N_{tot}$. Explicit equations for the Akaike information criterion, for $\kappa_{Cep}$, and for $\Delta\kappa_{Cep}$ are provided in the **Supporting Information**.

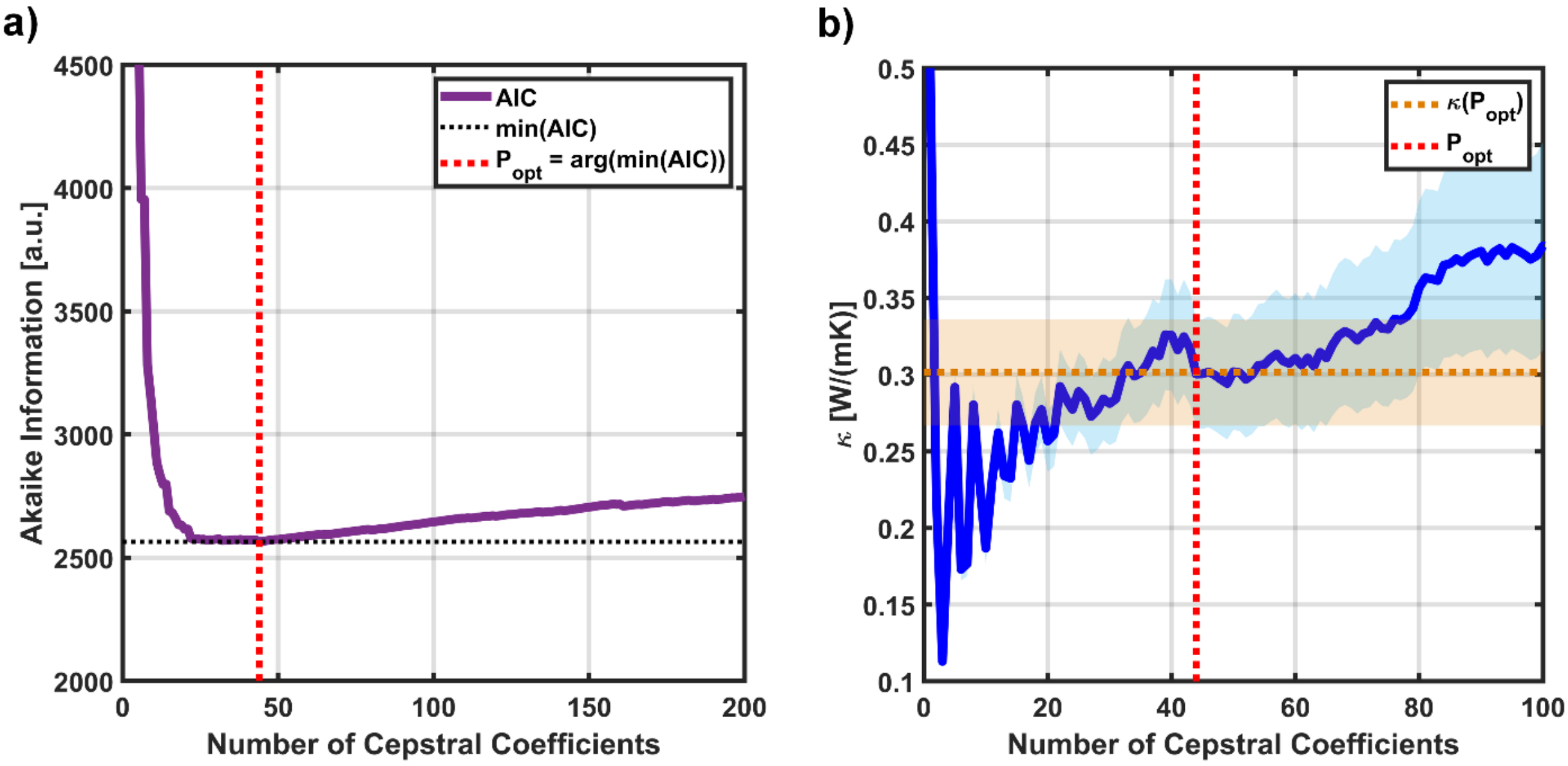

*Figure 6. Akaike information criterion $AIC_c$ (a) with global minimum (black dashed line) and optimal number of cepstral coefficients (red dashed line). (b) Corresponding thermal conductivity $\kappa_{Cep}$ as a function of cepstral coefficients (blue) with statistical uncertainties from the cepstral analysis (blue shaded region). The red dashed line marks the κ value for the optimal number of cepstral coefficients and the orange shaded horizontal region marks the uncertainty of the model averaged κ.*

To conclude this section, the above considerations are applied to the aforementioned data for MOF-5. Starting from the power spectrum in Figure 5, the cepstral coefficients, $C_n$, are calculated. For that $S(\omega_k)$ is considered only up to a certain cut-off frequency, F*, as following equation (9), only the low-frequency region of the power spectrum is of interest for the calculation of the cepstral coefficients. F* is the only non-deterministic parameter that needs to be chosen in the portrayed procedure. According to Ercole et al., it should be selected such that it incorporates the first prominent feature of the power spectrum.[32] Here, we set F* to 2.5 THz, thereby considering the part of $S(\omega_k)$ shown in Figure 5. In this way the prominent peak of $S(\omega_k)$ centered at 1.45 THz is included.  The robustness of the obtained value of κ with respect to the choice of F* is high, as discussed in the **Supporting Information**.

Figure 6a depicts $\kappa_{Cep}$ obtained after applying the above-described procedure as a function of the number P of cepstral coefficients used to reconstruct the power spectrum. Figure 6a, also the statistical uncertainty estimates as a region shaded in light blue. Its relatively large width is a consequence of using only a 1 ns trajectory, chosen here to allow a direct comparison with the data shown for the conventional GK analysis presented in Section 3.2. In line with equation (13) the uncertainty $\Delta\kappa_{Cep}$ increases when more cepstral coefficients are used to reconstruct $S(\omega_k)$. As outlined before, in order to pick the optimal pair ($\kappa_{Cep}$, $\Delta\kappa_{Cep}$) the Akaike information criterion $AIC_c$ is used,[84] which for the discussed MOF-5 case is depicted in Figure 6b. As can be seen in Figure 6b, the $AIC_c$ can exhibit several shallow local minima. To avoid an ambiguous selection of P*, we employ so called model averaging by assigning weights to possible local minima according to relative $AIC_c$ values.[84] The final conductivity κ and its error estimate are then obtained as the weighted average of the corresponding cepstral estimates $\kappa_{Cep}$. Further details on model averaging are provided in the **Supplementary Material**. As for the example case of MOF-5, the uncertainty region that takes model averaging into account is visualized in Figure 6b as a horizontal orange shaded region

# 4. Results and Discussion

The previous sections describe the working principle for standard GK simulations and for simulations augmented by cepstral analysis; in the following, the impact of various ambiguous process parameters on the results shall be discussed for the MOF-5 case considering trajectories with a total simulation time of 10 ns. Finally, also the situations for HKUST-1 and for ZIF-8 will be analyzed.

## 4.1. Time convergence of the Green-Kubo simulations

High thermal conductivity materials are characterized by slowly decaying HFACFs, and, hence, require long simulation times in order to capture the full correlation.[26] In contrast, as already discussed in section 3.2, the HFACF of MOFs decays very quickly due to their extremely low thermal conductivities.[27] At first sight this might appear as an indicator for requiring only moderate overall simulation times. However, in systems like MOFs the contributing portion of the HFACF is quickly buried in statistical noise. Consequently, the analysis time window $t_{seg}$, can be comparably small, but it might be useful to average over a significant number of sections of the trajectories. In order to investigate the convergence of the value of the thermal conductivity with the total simulation time, we proceed as follows: for a specific analysis time window $t_{seg}$, all ten trajectories with a total available simulation time of $T_{sim}$ = 10 ns are subdivided into $M= \lfloor T_{sim} / t_{seg} \rfloor$ segments. For the first segment of length [0, $t_{seg}$], we apply the procedures described above for the conventional approach and also include cepstral analysis. This results in a first $\kappa$ value after $t_{seg}$ simulation time for each of the approaches. Then, the same procedure is applied to the time interval [$t_{seg}$, $2 \cdot t_{seg}$], yielding another HFACF. After averaging over the two obtained HFACFs, again $\kappa$ values are extracted for “standard” and “cepstral” GK, resulting in the data points after a total of $2 \cdot t_{seg}$ simulation time. This process is repeated M-times. When setting $t_{seg}$ to 100 ps, one obtains the evolution of the thermal conductivity with simulation time shown in Figure 7. For the classical result we applied the same “freely” chosen parameters used already in section 3.2: a smoothing window for the $\kappa$ integral of 0.2 ps and an extraction point for $\kappa$ after 25 ps. For the cepstral analysis, the cutoff frequency was set to 2 THz. The total available simulation time of 10 ns is obtained by combining the ten available 1 ns trajectories in a random order. For the original graph referred to as “standard sequence” we combined the trajectories with their identifiers in ascending order. To illustrate the statistical nature of the process, we analyzed two additional random sequences of the ten 1 ns trajectories, which are plotted in green and blue. The data in Figure 7 convey three messages: (i) For short simulation times, the standard GK simulations produce essentially random (i.e., comparably large positive and also negative) values of the thermal conductivity, while the “GK+cepstral analysis” values remain at least in the range of the converged result, even when considering only a single 100 ps segment. (ii) For the latter simulations, the “onset” of convergence is reached for two of the trajectory orders already after 1 ns, and the thermal conductivity appears “safely converged” after ca. 5 ns. Conversely, there is a huge spread of the κ-vales after 1 ns for the standard GK simulations; a “reasonable convergence” appears to be reached after 5 ns; but for the entire duration of the simulations, rather significant fluctuations of κ occur. (iii) Finally, the “standard GK” and the “GK + cepstral analysis” values converge to two different thermal conductivities (0.36 W $m^{-1}$ $K^{-1}$ and 0.31 W $m^{-1}$ $K^{-1}$, respectively). We attribute this to the significant impact of the choice of ambiguous ad hoc parameters on the results in the “standard GK” simulations, as will become apparent from the next section. Notably, the value obtained when including cepstral analysis is exceptionally close to the experimental value of 0.32 W $m^{-1}$ $K^{-1}$. It should, however, be mentioned that it cannot be excluded that this value is affected by scattering at imperfections, as can be inferred from the vanishing temperature dependence of the thermal conductivity of MOF-5.[12]

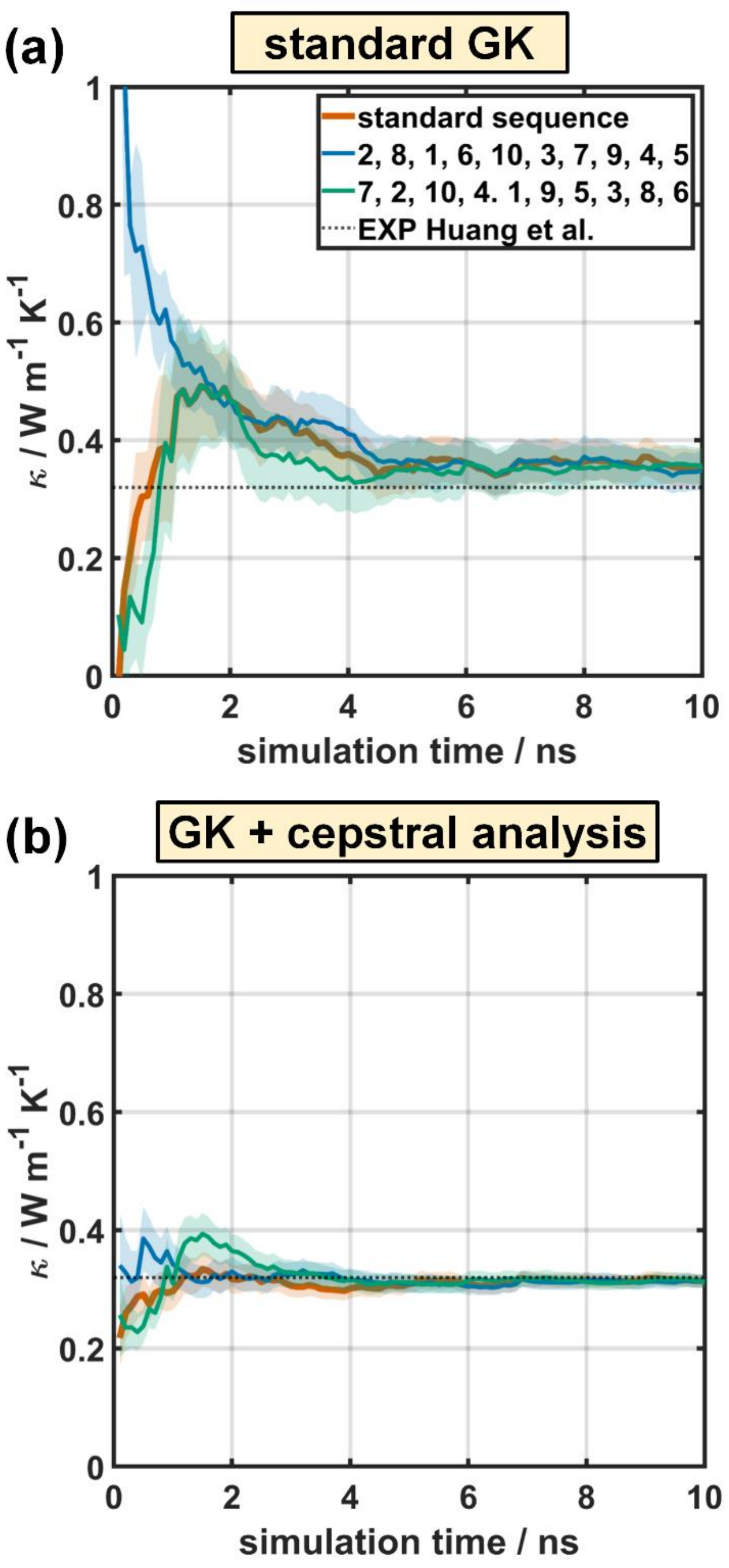

*Figure 7. Time convergence behavior of the calculated thermal conductivity of MOF-5 obtained from conventional GK analysis (a) and from GK combined with cepstral analysis (b). In each panel, the three curves correspond to different permutations of the same set of ten independent 1 ns simulations, where the standard ordering (corresponding to the actual sequence of our simulations) is shown in red. Shaded regions indicate statistical uncertainties. For the conventional GK approach in (a), the standard error of the mean is shown, whereas for the*

*cepstral analysis in (b), uncertainties are obtained as described in Section 3.3. The experimental value (dotted line) was taken from Huang et al.*[12]

## 4.2. Many knobs to turn – Influence of ad-hoc simulation parameters on Green-Kubo simulations

Another complication with "classical" GK simulations besides the more gradual convergence discussed above, is the need for choosing several ad-hoc parameters when analyzing the HFACF (see discussion in section 3.2). Thus, in the following, possible risks and pitfalls arising from an unfortunate choice of these parameters are discussed. The considered ad hoc parameters are the analysis window $t_{seg}$, the width of the smoothing window, when determining κ from integrating the HFACF, see Figure 4a, as well as the extraction point for $\kappa$. The smoothing window for calculating the HFACF is fixed at 0.5 ps as changing the width of that window and the one for the calculation of κ are expected to impact results similarly. First, the impact of the extraction point of κ and the width of the associated smoothing window will be discussed for a fixed analysis time of 100 ps. As a second step, the impact of the choice of the analysis time, $t_{seg}$, will be studied.

Figure 8a shows the HFACF for MOF-5 together with the thermal conductivity obtained from the Green-Kubo integral (c.f., equation (8)). Notice the depicted quantities in Figure 8a, are obtained from a 1 ns trajectory, hence 10 times averaged. Additionally, different possible extraction points, $\tau_{ext}$, for the thermal conductivity are shown. For the current test, they were chosen to be 15, 25, 40 and 60 ps respectively. For determining the evolution of κ with the simulation time, the respective values of κ are again averaged over an increasing number of 100 ps trajectories. Compared to Figure 7, only the "standard sequence" of the ten 1 ns trajectories is considered in Figure 8 for the sake of visibility.

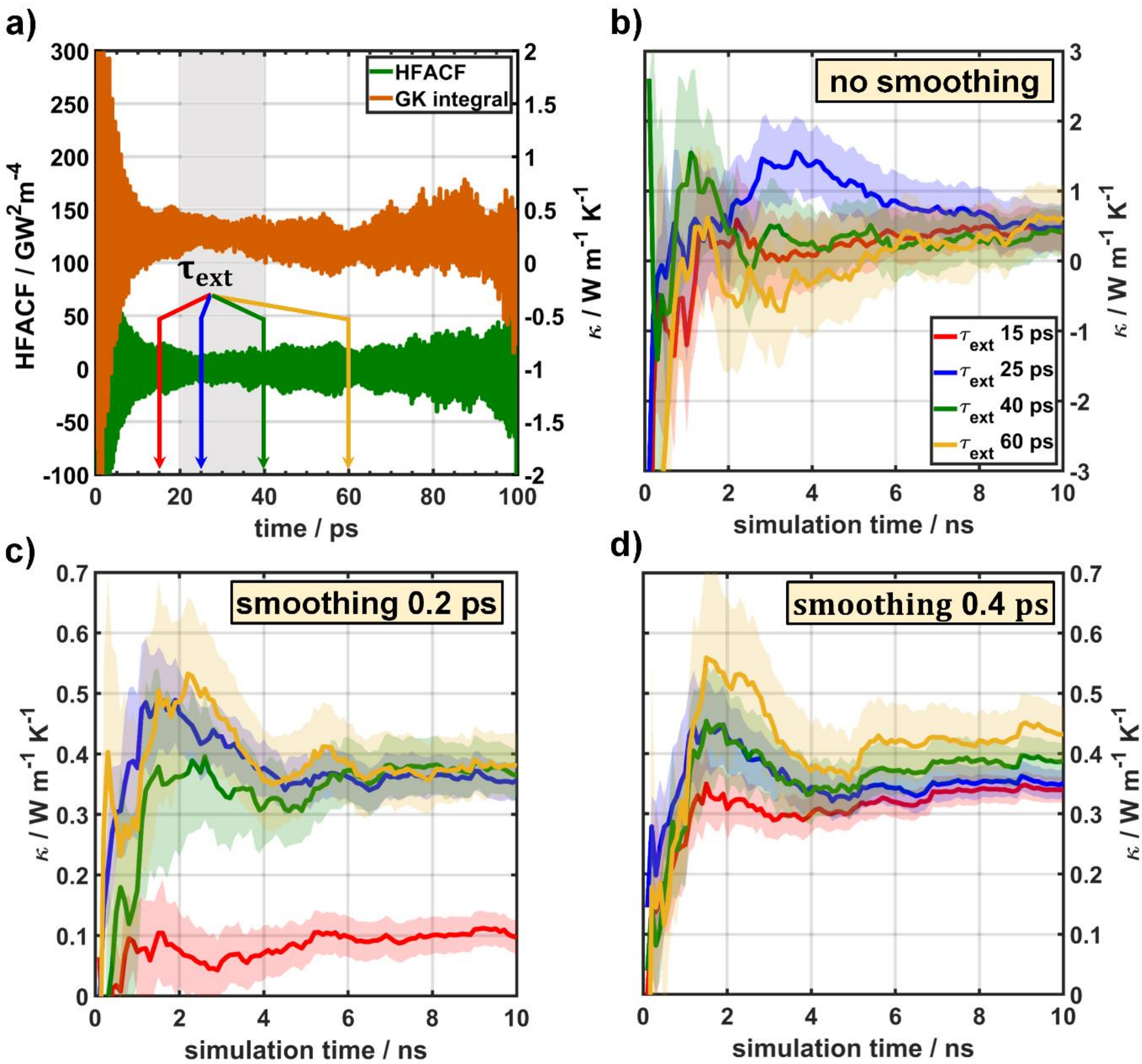

*Figure 8. Analysis of the influence of the extraction point and the κ-smoothing window width on the thermal conductivity obtained from the "standard GK" method. (a) Exemplary HFACF and corresponding κ value for an analysis window of 100 ps for the first calculated trajectory. The extraction times, $\tau_{ext}$, used in the subsequent panels are indicated by arrows of corresponding colors. The gray shaded region highlights the apparent plateau region of the integral. (b) Time convergence behavior of the thermal conductivity without running-average smoothing of κ. (c) Same analysis as in (b), but using a running average smoothing window of width 0.2 ps; (d) same analysis as in (c), but with a smoothing window width of 0.4 ps. The shaded regions show the standard uncertainty of the mean κ value.*

As shown in Figure 8b, in the extreme case, when no running average smoothing is applied, $\kappa$ oscillates erratically for the first few ns, adopting also physically meaningless large negative values. Only when averaging over trajectories with a total simulation time > 6 ns the $\kappa$-values gradually start making sense at least qualitatively. Nevertheless, even for the full 10 ns simulation time, values ranging from 0.40 W $m^{-1}$ $K^{-1}$ to 0.63 W $m^{-1}$ $K^{-1}$ are observed and the situation remains far from converged (mind the strongly increased $\kappa$-scale in panel (b)). The situation improves when applying a running average with a window of 0.2 ps (see Figure 8c). In that case, at least for three of the chosen extraction points, the spread of the $\kappa$-values range between 0.35 W $m^{-1}$ $K^{-1}$ and 0.38 W $m^{-1}$ $K^{-1}$ and convergence appears to be reached beyond 7 ns. The trend for the fourth extraction point at 15 ps is, however, fundamentally different, yielding a thermal conductivity of only ~0.1 W $m^{-1}$ $K^{-1}$. This is attributed to a small value of $\tau_{ext}$ such that significant portions of the GK integral are missed. When only somewhat increasing the averaging window to 0.4 ps, the situation changes fundamentally: while convergence appears to occur at similar simulation times as in the previous case, now there is again a significant spread in the thermal conductivities after 10 ns with only the 10 ps and the 25 ps values being seemingly consistent.

As argued in section 3.2, for most of the MD trajectories obtained for MOF-5 in our simulations, the supposedly ideal extraction point amounts to 25 ps. This extraction point at least for long enough trajectories also does not display any pathological behavior in the evolutions shown in Figure 8. Thus, in Figure 9, the corresponding convergence plots for varying averaging windows at an extraction point of 25 ps are compared. Apart from the curve without smoothing, consistent trends are obtained for smoothing widths of 0.2, 0.4 and 10 ps with converged results obtained after ca. 5 ns.

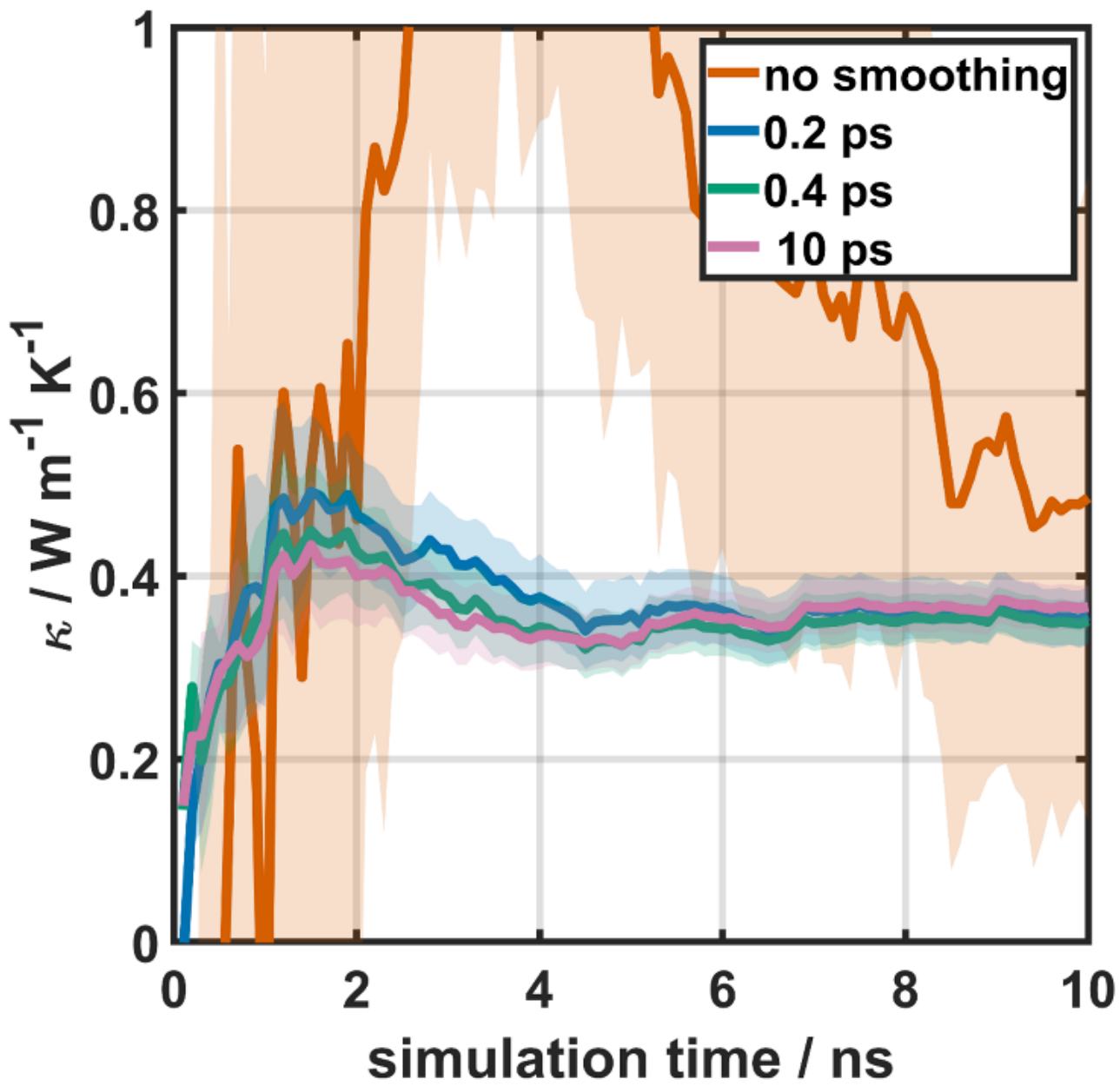

*Figure 9. Comparison of the time convergence behavior obtained without any smoothing of the evolution of $\kappa$ and using running-average smoothing window widths of 0.2 ps, 0.4 ps, and 10 ps for a fixed extraction time of 25 ps.*

The above-discussed data suggest a highly erratic convergence behavior in classical GK simulations, especially regarding the choice of the extraction point. I.e., identifying suitable regions in the cumulative GK integral (8) appears to be crucial for obtaining meaningful transport coefficients, but this is a non-trivial problem due to the integral's random-walk type characteristics. Unfortunately, the choice of optimal parameters for classical GK simulations is expected to be highly system-specific, which renders the application of the classical GK approach for a high-throughput screening of MOFs challenging.[27] In passing, we note that the only ad hoc parameter in GK simulations combined with cepstral analysis is the cut-off frequency chosen for the spectral function when calculating the cepstrum. In that context we show in the **Supporting Information** that the choice of F* only has a minor impact on the eventually calculated thermal conductivity provided that one follows the guideline from the paper by Ercole et al.[32] and chooses F* such that the first major feature of the spectral function is contained in the analysis.

This being said, there is yet another parameter that appears both in classical GK simulations as well as when employing cepstral analysis This is magnitude of the analysis time window, $t_{seg}$, i.e., the actual sections into which the overall trajectories are cut. These are the time windows for which the whole GK analysis (with or without cepstral analysis) is performed and over which the averaging occurs yielding the convergence plots. In section 3, $t_{seg}$ has been set to 1 ns, while in section 4 it so far has been 100 ps.

The impact of the choice of $t_{seg}$ on the results for MOF-5 is illustrated in Figures 10a and b. The solid lines describe the situation including cepstral analysis (with a universal frequency cut-off 2.5 THz). The curves (which are identical in panels (a) and (b)) essentially lie on top of each other, showing that the choice of $t_{seg}$ has even less impact than changing the order in which the 1 ns simulation segments are stitched together (compare Figure 7). Moreover, the solid lines in Figure 10a and b testify to the rapid convergence, when including cepstral analysis. The situation is fundamentally different in "standard GK" simulations: here, we compare two cases, namely setting the extraction point to a fixed fraction (here 20%) of $t_{seg}$ (panel (a)) and choosing a fixed extraction point (the 25 ns used also before; panel (b)). In both instances pronounced variations over the course of the simulation time are observed. Overall, the final values are, however, still at least in the range of the values obtained including cepstral analysis and especially for the fixed extraction time of 25 ps in panel (b) the "standard" GK simulations converge to similar thermal conductivities of ~0.36 W $m^{-1}$ $K^{-1}$, which is by approximately 16 % higher than the value obtained when including cepstral analysis (0.31 $Wm^{-1}K^{-1}$).

## 4.3. Robustness of the Cepstral approach – Three prototypical MOFs

This raises the question of how the different approaches fare when applying them to other materials (albeit keeping most of the simulation parameters at the values that so far worked reasonably well for MOF-5). The two additional materials considered in this section are HKUST-1 and ZIF-8, which have already been introduced in section 1.1 GK simulations combined with cepstral analysis (at the aforementioned universal frequency cut-off of 2.5 THz) again provide perfectly consistent and fast-converging results as shown by the solid lines in Figures 10c and e (replicated in panels (d) and (f)). In contrast, the "standard GK" simulations face a number of problems: For HKUST-1 in panel (c) only the simulation with $t_{seg}$=500 ps (and, thus, an extraction time of 100 ps) provides a meaningful result. When choosing $t_{seg}$=100 ps, the extraction time becomes 20 ps, which is apparently too short for HKUST-1, whose thermal conductivity is approximately twice as large as that of MOF-5; i.e., here one appears to run into problems similar to those encountered when modelling MOF-5 with an extraction time of 15 ps and a smearing window of 0.2 ps in Figure 8c. Interestingly, while the extraction time of 20 ps produces a large negative thermal conductivity in Figure 10c, a slightly larger value of 0.25 ps in Figure 10d results in excessively large positive values. This suggests that the corresponding extraction times are in a region of significant noise in the GK integral. Similarly, strong variations of the largely inconsistent "standard" GK results are found also for ZIF-8 in Figure 10e and f.

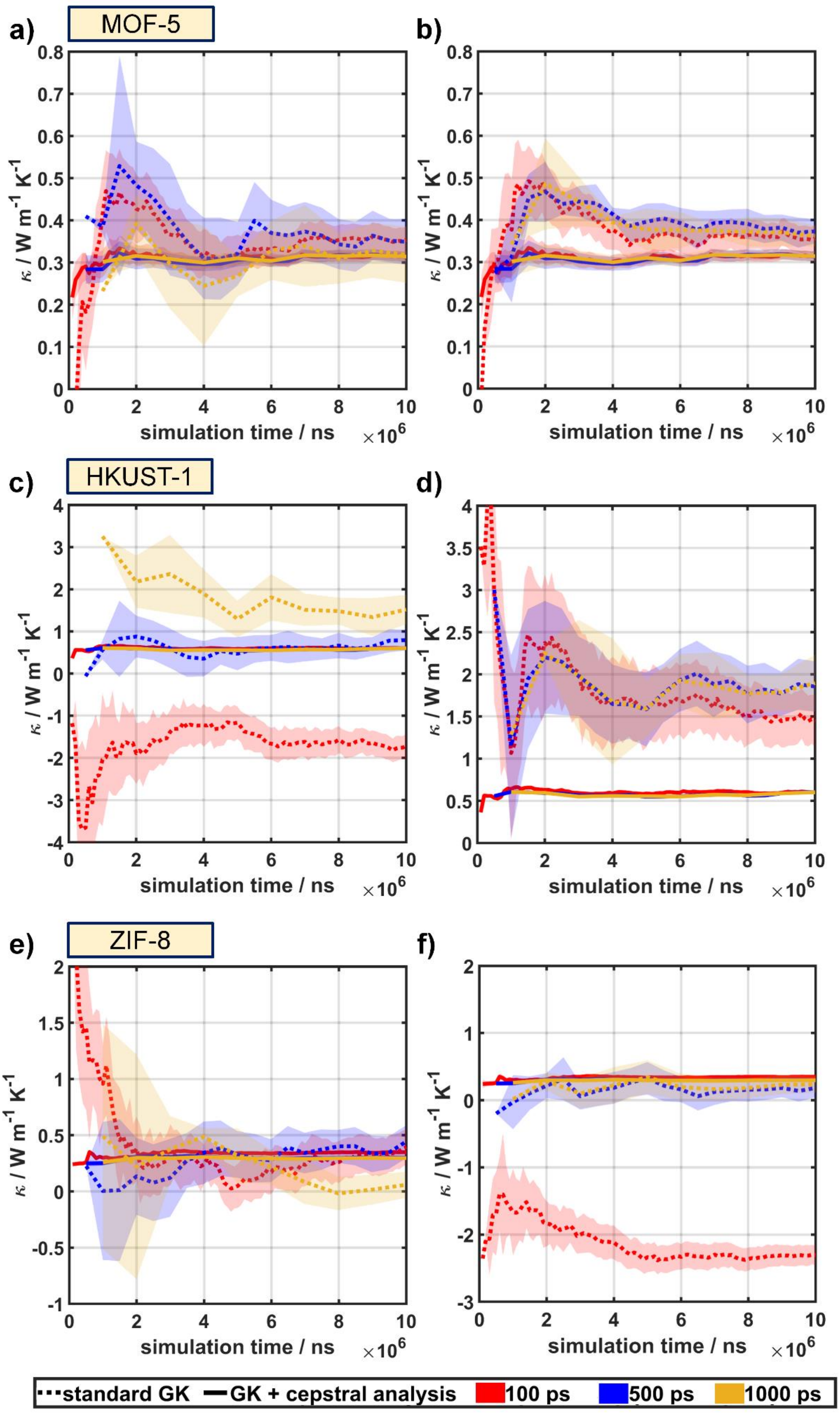
a)
MOF-5
b)
c)
HKUST-1
d)
e)
ZIF-8
f)
κ / W m-1 K-1
simulation time / ns
×10^6
standard GK
GK + cepstral analysis
100 ps
500 ps
1000 ps

*Figure 10. Comparison of thermal conductivity convergence behavior for MOF-5 ((a),(b)), HKUST-1 ((c),(d)), and ZIF-8 ((e),(f)) obtained using different analysis window lengths (100 ps, 500 ps, and 1000 ps). Solid lines correspond to conventional GK integration, whereas dotted lines denote GK combined with cepstral analysis. In the left column ((a),(c),(e)), the extraction time is chosen as one fifth of the corresponding analysis window length. In the right column ((b),(d),(f)), a fixed extraction time of 25 ps is used for all analysis windows. For the "standard" GK simulations, a smearing window of 0.2 ps was chosen for the GK integral and when including cepstral analysis, the cut-off frequency was set to 2.5 THz.*

Using a total simulation time of 10 ns and a correlation length of 500 ps, the GK approach combined with cepstral analysis yields thermal conductivities of **0.31 W m$^{-1}$ K$^{-1}$** for MOF-5, **0.60 W m$^{-1}$ K$^{-1}$** for HKUST-1, and **0.30 W m$^{-1}$ K$^{-1}$** for ZIF-8. In case of MOF-5 and HKUST-1, these results show excellent agreement with available single-crystal experimental data: Huang et al.[13] report 0.32 W m$^{-1}$ K$^{-1}$ for MOF-5 employing the steady state heat flow method, and Babaei et al. [14] report (0.69 ± 0.05) W m$^{-1}$ K$^{-1}$ for HKUST-1 using the frequency-domain thermoreflectance approach. This corresponds to deviations of approximately 10 %. For comparison, Ying *et al.* reported thermal conductivities of 0.46 W m$^{-1}$ K$^{-1}$ for MOF-5, 0.78 W m$^{-1}$ K$^{-1}$ for HKUST-1, and 0.57 W m$^{-1}$ K$^{-1}$ for ZIF-8 using supercells of the same size as those employed here.[25] The simulated values are also consistent with the results we obtained using similar machine learned potentials when employing anharmonic lattice dynamics (0.36 W m$^{-1}$K$^{-1}$ for MOF-5 and 0.60 Wm$^{-1}$K$^{-1}$ for HKUST-1) and for MOF-5 also for the case of nonequilibrium molecular dynamics based techniques (0.32W m$^{-1}$ K$^{-1}$).[60] For ZIF-8, the cepstral estimate of (0.30 ± 0.01) W m$^{-1}$ K$^{-1}$ is roughly a factor of two lower than the experimental value of (0.64 ± 0.09) W m$^{-1}$ K$^{-1}$ reported by Zhang et al.[15] using the Raman-Resistance Temperature Detector method.[63] It seems that the $\kappa$ value we determined for ZIF-8 is closer to thin film measurements: Using the 3ω method, Cui et al., report a thermal conductivity of (0.326 ± 0.002) W $m^{-1}$ $K^{-1}$ for ZIF-8 thin films.[85]

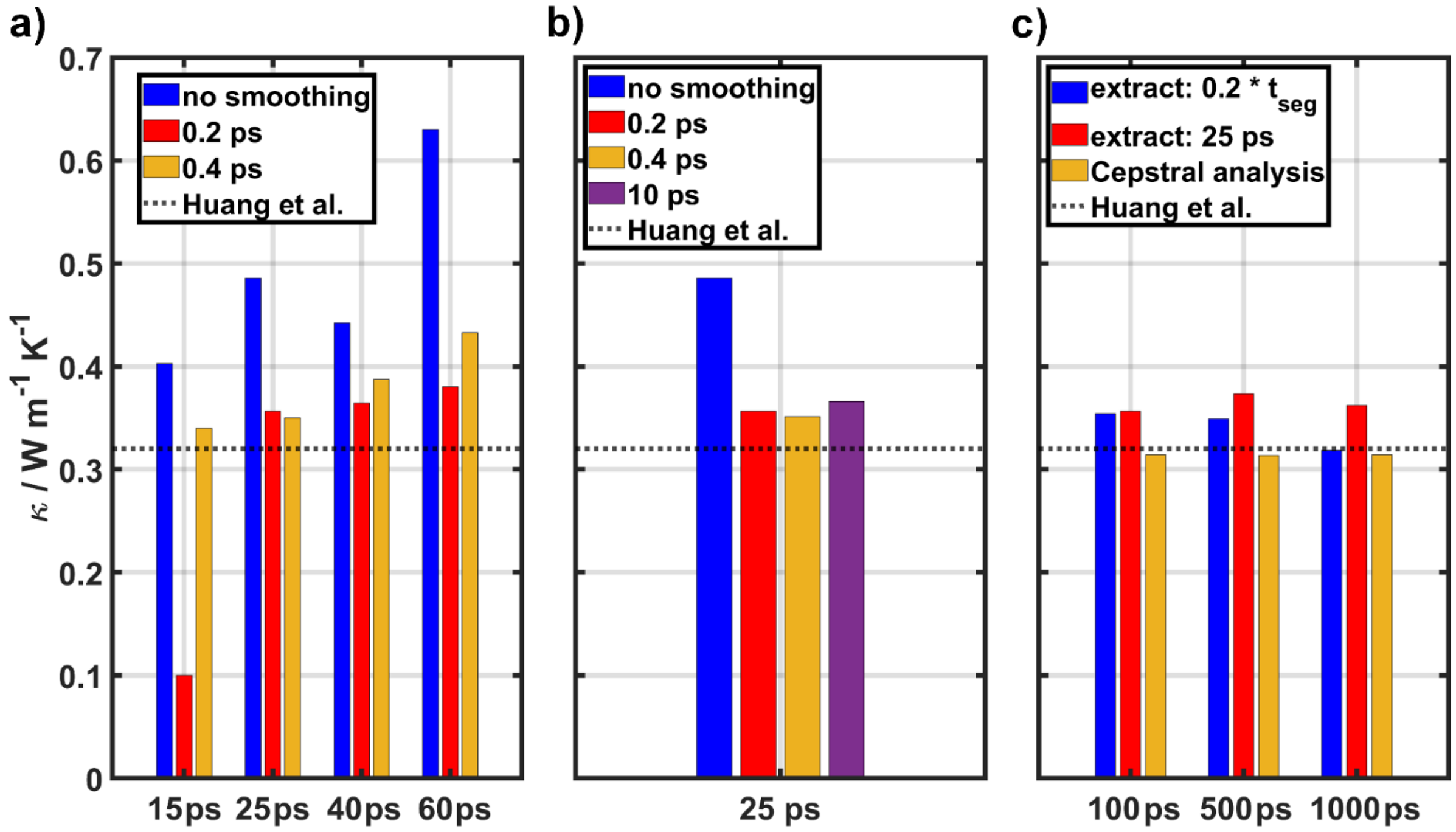

*Figure 11. Summary of thermal conductivity values for MOF-5 obtained after a simulation time of 10 ns, based on the analyses discussed in Figures 8–10. The experimental single crystal value reported by Huang et al. is shown as a black dashed line. (a) Dependence of the direct GK thermal conductivity on extraction point and running-average smoothing window. x-axis labels denote corresponding extraction points. (b) Dependence on smoothing-window width for a fixed extraction point of 25 ps. (c) Dependence on time segment size $t_{seg}$ for direct GK analysis using two extraction strategies (fixed 25 ps and 0.2 $t_{seg}$), compared to cepstral analysis.*

# 5. Conclusion

In this work, we assessed the applicability of cepstral analysis for reducing the impact of noise when evaluating GK integrals for determining thermal conductivities in MOFs. We explicitly demonstrated how user defined ad-hoc parameters introduced along the simulation workflow of “standard GK” simulations can massively influence the obtained results. Smoothing choices, correlation length selection, and subjective extraction criteria can lead to erratic convergence and substantial spread in direct GK estimates – limitations that become particularly problematic when reproducibility and automation are required. This is illustrated in Figure 11, where panel (a) compares the results of “standard GK” simulations for MOF-5 for different soothing windows and extraction points (for details see Figure 8 and section 4.1); panel (b) contains “standard GK” results for different smoothing windows and fixed extraction points. Finally, panel (c) shows the dependence of the obtained results on the analysis time window for both the standard GK approach and the GK+cepstral method.

A particular technical advantage of applying cepstral analysis in this context is that much fewer ambiguously user defined parameters are required than for “standard GK”, as summarized in Table 1.

*Table 1. Overview of standard ad-hoc parameter choices during GK simulations of thermal conductivities.*

| Parameter | Direct - GK | Cepstral Analysis - GK |
|---|---|---|
| **correlation length $t_{corr}$** | → influences formation of plateau region in cumulative $\kappa$-integral<br><br>→ needs manual inspection<br>✗ | → robust against specific selections of $t_{corr}$<br><br>✓ |
| **extraction point $\tau_{ext}$** | → requires ad-hoc (mostly by inspection by eye) identification of suitable plateau regions<br><br>→ needs manual inspection<br>✗ | → no extraction points, $\kappa$ is obtained from low frequency limit of the power spectrum $S(0)$ directly<br><br>✓ |
| **smoothing window** | → can introduce significant bias for unfortunate selection of extraction points and correlation lengths<br><br>→ needs manual inspection<br>✗ | → method does not rely on any moving average smoothing<br><br>✓ |

A further general observation is that the robust thermal conductivity values that also agree well with experiments for the three considered systems are obtained at comparably short simulation times of only a few ns when exploiting cepstral analysis.

Nevertheless, a complication that can be encountered when employing cepstral analysis needs to be mentioned: as the thermal conductivity is calculated from the value of the spectral function at $\omega \to 0$, simulations on systems with significantly larger thermal conductivities than MOFs (for which the spectral function rise steeply close to $\omega$=0) might again run into numerical instabilities. For classical MOFs or other low thermal conductivity materials like organic semiconductors, cepstral analysis is an extremely useful tool for boosting the accuracy of GK-based heat transport simulations.

**Acknowledgements**

We acknowledge the Graz University of Technology for financial support through the Lead Project Porous Materials @ Work for Sustainability (LP-03). The research was funded in part by the Austrian Science Fund (FWF) [https://doi.org/10.55776/P33903, https://doi.org/10.55776/P36129 and https://doi.org/10.55776/COE5]. For the purpose of open access, the authors have applied a CC-BY public copyright license to any author accepted manuscript version arising from this submission. Furthermore, the authors acknowledge that Google Gemini was used to support the literature search.

1. Moghadam, P. Z. *et al.* Development of a Cambridge Structural Database Subset: A Collection of Metal–Organic Frameworks for Past, Present, and Future. *Chem. Mater.* **29**, 2618–2625 (2017).
2. Li, T. *et al.* Scalable and efficient solar-driven atmospheric water harvesting enabled by bidirectionally aligned and hierarchically structured nanocomposites. *Nat. Water* **1**, 971–981 (2023).
3. Banlusan, K. Shock wave energy absorption via structural phase transition and bond breakage in metal–organic frameworks. *J. Chem. Phys.* **163**, 014703 (2025).
4. Sun, Y. *et al.* High-rate nanofluidic energy absorption in porous zeolitic frameworks. *Nat. Mater.* **20**, 1015–1023 (2021).
5. Shupletsov, L. *et al.* Linker Conformation Controls Oxidation Potentials and Electrochromism in Highly Stable Zr-Based Metal–Organic Frameworks. *J. Am. Chem. Soc.* **146**, 25477–25489 (2024).
6. Xu, X. *et al.* Iontronics Using V2CTx MXene-Derived Metal–Organic Framework Solid Electrolytes. *ACS Nano* **14**, 9840–9847 (2020).
7. Senkovska, I., Bon, V., Mosberger, A., Wang, Y. & Kaskel, S. Adsorption and Separation by Flexible MOFs. *Adv. Mater.* **37**, 2414724 (2025).
8. Lin, J.-B. *et al.* A scalable metal-organic framework as a durable physisorbent for carbon dioxide capture. *Science* **374**, 1464–1469 (2021).
9. Wieme, J. *et al.* Thermal Engineering of Metal–Organic Frameworks for Adsorption Applications: A Molecular Simulation Perspective. *ACS Appl. Mater. Interfaces* **11**, 38697–38707 (2019).
10. Beckner, M. & Dailly, A. A pilot study of activated carbon and metal–organic frameworks for methane storage. *Appl. Energy* **162**, 506–514 (2016).
11. Babaei, H., McGaughey, A. J. H. & Wilmer, C. E. Transient Mass and Thermal Transport during Methane Adsorption into the Metal–Organic Framework HKUST-1. *ACS Appl. Mater. Interfaces* **10**, 2400–2406 (2018).
12. Huang, B. L. *et al.* Thermal conductivity of a metal-organic framework (MOF-5): Part II. Measurement. *Int. J. Heat Mass Transf.* **50**, 405–411 (2007).

13. Babaei, H. *et al.* Observation of reduced thermal conductivity in a metal-organic framework due to the presence of adsorbates. *Nat. Commun.* **11**, 4010 (2020).

14. Huang, J., Fan, A., Xia, X., Li, S. & Zhang, X. In Situ Thermal Conductivity Measurement of Single-Crystal Zeolitic Imidazolate Framework-8 by Raman-Resistance Temperature Detectors Method. *ACS Nano* **14**, 14100–14107 (2020).

15. Wieser, S. *et al.* Identifying the Bottleneck for Heat Transport in Metal–Organic Frameworks. *Adv. Theory Simul.* **4**, 2000211 (2021).

16. Müller-Plathe, F. A simple nonequilibrium molecular dynamics method for calculating the thermal conductivity. *J. Chem. Phys.* **106**, 6082–6085 (1997).

17. Li, Z. *et al.* Influence of thermostatting on nonequilibrium molecular dynamics simulations of heat conduction in solids. *J. Chem. Phys.* **151**, 234105 (2019).

18. Sellan, D. P., Landry, E. S., Turney, J. E., McGaughey, A. J. H. & Amon, C. H. Size effects in molecular dynamics thermal conductivity predictions. *Phys. Rev. B* **81**, 214305 (2010).

19. Ferreira de Souza, N., Mercier Franco, L. F. & Coasne, B. Consistency of Equilibrium and Nonequilibrium Molecular Dynamics to Assess Thermal Conductivity. *J. Chem. Eng. Data* **71**, 1022–1032 (2026).

20. Vercouter, A., Lemaur, V., Melis, C. & Cornil, J. Computing the Lattice Thermal Conductivity of Small-Molecule Organic Semiconductors: A Systematic Comparison of Molecular Dynamics Based Methods. *Adv. Theory Simul.* **6**, 2200892 (2023).

21. Baroni, S. Green and Kubo forge the arrow of time. *Mol. Phys.* **123**, e2388300 (2025).

22. Knoop, F., Scheffler, M. & Carbogno, C. Ab initio Green-Kubo simulations of heat transport in solids: Method and implementation. *Phys. Rev. B* **107**, 224304 (2023).

23. Kubo, R. Statistical-Mechanical Theory of Irreversible Processes. I. General Theory and Simple Applications to Magnetic and Conduction Problems. *J. Phys. Soc. Jpn.* **12**, 570–586 (1957).

24. Green, M. S. Markoff Random Processes and the Statistical Mechanics of Time-Dependent Phenomena. *J. Chem. Phys.* **20**, 1281–1295 (1952).

25. Ying, P. *et al.* Sub-Micrometer Phonon Mean Free Paths in Metal–Organic Frameworks Revealed by Machine Learning Molecular Dynamics Simulations. *ACS Appl. Mater. Interfaces* **15**, 36412–36422 (2023).

26. Wieser, S., Cen, Y.-J., Madsen, G. K. H. & Carrete, J. Accelerating First-Principles Molecular-Dynamics Thermal Conductivity Calculations for Complex Systems. *J. Chem. Theory Comput.* **22**, 513–527 (2026).

27. Islamov, M. *et al.* High-throughput screening of hypothetical metal-organic frameworks for thermal conductivity. *Npj Comput. Mater.* **9**, 11 (2023).

28. Oliveira, L. de S. & Greaney, P. A. Method to manage integration error in the Green-Kubo method. *Phys. Rev. E* **95**, 023308 (2017).

29. Marcolongo, A., Ercole, L. & Baroni, S. Gauge Fixing for Heat-Transport Simulations. *J. Chem. Theory Comput.* **16**, 3352–3362 (2020).

30. Otero-Lema, M. *et al.* KUTE: Green–Kubo Uncertainty-Based Transport Coefficient Estimator. *J. Chem. Inf. Model.* **65**, 3477–3487 (2025).

31. Childers, D. G., Skinner, D. P. & Kemerait, R. C. The cepstrum: A guide to processing. *Proc. IEEE* **65**, 1428–1443 (1977).

32. Ercole, L., Marcolongo, A. & Baroni, S. Accurate thermal conductivities from optimally short molecular dynamics simulations. *Sci. Rep.* **7**, 15835 (2017).

33. Ercole, L., Bertossa, R., Bisacchi, S. & Baroni, S. SporTran: A code to estimate transport coefficients from the cepstral analysis of (multivariate) current time series. *Comput. Phys. Commun.* **280**, 108470 (2022).

34. Bertossa, R., Grasselli, F., Ercole, L. & Baroni, S. Theory and Numerical Simulation of Heat Transport in Multicomponent Systems. *Phys. Rev. Lett.* **122**, 255901 (2019).

35. Pegolo, P., Drigo, E., Grasselli, F. & Baroni, S. Transport coefficients from equilibrium molecular dynamics. *J. Chem. Phys.* **162**, 064111 (2025).

36. Novikov, I. S., Gubaev, K., Podryabinkin, E. V. & Shapeev, A. V. The MLIP package: moment tensor potentials with MPI and active learning. *Mach. Learn. Sci. Technol.* **2**, 025002 (2020).

37. Shapeev, A. V. Moment Tensor Potentials: A Class of Systematically Improvable Interatomic Potentials. *Multiscale Model. Simul.* **14**, 1153–1173 (2016).

38. Li, H., Eddaoudi, M., O'Keeffe, M. & Yaghi, O. M. Design and synthesis of an exceptionally stable and highly porous metal-organic framework. *Nature* **402**, 276–279 (1999).

39. Chui, S. S.-Y., Lo, S. M.-F., Charmant, J. P. H., Orpen, A. G. & Williams, I. D. A Chemically Functionalizable Nanoporous Material [Cu3(TMA)2(H2O)3]n. *Science* **283**, 1148–1150 (1999).

40. Park, K. S. *et al.* Exceptional chemical and thermal stability of zeolitic imidazolate frameworks. *Proc. Natl. Acad. Sci.* **103**, 10186–10191 (2006).

41. Huang, B. L., McGaughey, A. J. H. & Kaviany, M. Thermal conductivity of metal-organic framework 5 (MOF-5): Part I. Molecular dynamics simulations. *Int. J. Heat Mass Transf.* **50**, 393–404 (2007).

42. Zhang, S., Liu, J. & Liu, L. Insights into the thermal conductivity of MOF-5 from first principles. *RSC Adv.* **11**, 36928–36933 (2021).

43. Effect of pore size and shape on the thermal conductivity of metal-organic frameworks. *Chem. Sci.* **8**, 583–589 (2016).

44. Islamov, M., Babaei, H. & Wilmer, C. E. Influence of Missing Linker Defects on the Thermal Conductivity of Metal–Organic Framework HKUST-1. *ACS Appl. Mater. Interfaces* **12**, 56172–56177 (2020).

45. Fan, H., Li, Z. & Zhou, Y. Influence of adsorbed gas molecules on thermal transport in metal-organic framework HKUST-1. *Phys. Rev. Appl.* **23**, 054006 (2025).

46. Zhang, X. & Jiang, J. Thermal Conductivity of Zeolitic Imidazolate Framework-8: A Molecular Simulation Study. *J. Phys. Chem. C* **117**, 18441–18447 (2013).

47. Ying, P., Zhang, J., Zhang, X. & Zhong, Z. Impacts of Functional Group Substitution and Pressure on the Thermal Conductivity of ZIF-8. *J. Phys. Chem. C* **124**, 6274–6283 (2020).

48. Lock, N. *et al.* Elucidating Negative Thermal Expansion in MOF-5. *J. Phys. Chem. C* **114**, 16181–16186 (2010).

49. Alowasheeir, A. *et al.* Synthesis of millimeter-scale ZIF-8 single crystals and their reversible crystal structure changes. *Sci. Technol. Adv. Mater.* **25**, 2292485 (2024).

50. Lee, J. *et al.* Metal–organic framework materials as catalysts. *Chem. Soc. Rev.* **38**, 1450–1459 (2009).

51. Pascanu, V., González Miera, G., Inge, A. K. & Martín-Matute, B. Metal–Organic Frameworks as Catalysts for Organic Synthesis: A Critical Perspective. *J. Am. Chem. Soc.* **141**, 7223–7234 (2019).

52. Gangu, K. K., Maddila, S. & Jonnalagadda, S. B. The pioneering role of metal–organic framework-5 in ever-growing contemporary applications – a review. *RSC Adv.* **12**, 14282–14298 (2022).

53. Denning, S. *et al.* Metal–Organic Framework HKUST-1 Promotes Methane Hydrate Formation for Improved Gas Storage Capacity. *ACS Appl. Mater. Interfaces* **12**, 53510–53518 (2020).

54. Liang, W. *et al.* Metal–Organic Framework-Based Enzyme Biocomposites. *Chem. Rev.* **121**, 1077–1129 (2021).

55. Chafiq, M., Chaouiki, A., Suhartono, T. & Ko, Y. G. Albumin protein encapsulation into a ZIF-8 framework with Co-LDH-based hierarchical architectures for robust catalytic reduction. *J. Mater. Chem. A* **11**, 23984–23998 (2023).

56. Liang, K. *et al.* Biomimetic mineralization of metal-organic frameworks as protective coatings for biomacromolecules. *Nat. Commun.* **6**, 7240 (2015).

57. Kresse, G. & Hafner, J. Ab initio molecular dynamics for liquid metals. *Phys. Rev. B* **47**, 558–561 (1993).

58. Kresse, G. & Furthmüller, J. Efficiency of ab-initio total energy calculations for metals and semiconductors using a plane-wave basis set. *Comput. Mater. Sci.* **6**, 15–50 (1996).

59. Kresse, G. & Furthmüller, J. Efficient iterative schemes for ab initio total-energy calculations using a plane-wave basis set. *Phys. Rev. B* **54**, 11169–11186 (1996).

60. Wieser, S. & Zojer, E. Machine learned force-fields for an Ab-initio quality description of metal-organic frameworks. *Npj Comput. Mater.* **10**, 18 (2024).

61. Strasser, N., Wieser, S. & Zojer, E. Predicting Spin-Dependent Phonon Band Structures of HKUST-1 Using Density Functional Theory and Machine-Learned Interatomic Potentials. *Int. J. Mol. Sci.* **25**, 3023 (2024).

62. Perdew, J. P., Burke, K. & Ernzerhof, M. Generalized Gradient Approximation Made Simple. *Phys. Rev. Lett.* **77**, 3865–3868 (1996).

63. Perdew, J. P., Burke, K. & Ernzerhof, M. Generalized Gradient Approximation Made Simple [Phys. Rev. Lett. 77, 3865 (1996)]. *Phys. Rev. Lett.* **78**, 1396–1396 (1997).

64. Grimme, S., Antony, J., Ehrlich, S. & Krieg, H. A consistent and accurate ab initio parametrization of density functional dispersion correction (DFT-D) for the 94 elements H-Pu. *J. Chem. Phys.* **132**, 154104 (2010).

65. Grimme, S., Ehrlich, S. & Goerigk, L. Effect of the damping function in dispersion corrected density functional theory. *J. Comput. Chem.* **32**, 1456–1465 (2011).

66. Jinnouchi, R., Karsai, F. & Kresse, G. On-the-fly machine learning force field generation: Application to melting points. *Phys. Rev. B* **100**, 014105 (2019).

67. Jinnouchi, R., Karsai, F., Verdi, C., Asahi, R. & Kresse, G. Descriptors representing two- and three-body atomic distributions and their effects on the accuracy of machine-learned inter-atomic potentials. *J. Chem. Phys.* **152**, 234102 (2020).

68. Thompson, A. P. *et al.* LAMMPS - a flexible simulation tool for particle-based materials modeling at the atomic, meso, and continuum scales. *Comput. Phys. Commun.* **271**, 108171 (2022).

69. Hamakawa, T., McGaughey, A. J. H. & Shiomi, J. Accurate heat flux formula and thermal conductivity calculation in molecular dynamics simulations with machine learning potentials. *J. Appl. Phys.* **138**, 055105 (2025).

70. Tai, S. T., Wang, C., Cheng, R. & Chen, Y. Revisiting Many-Body Interaction Heat Current and Thermal Conductivity Calculations Using the Moment Tensor Potential/LAMMPS Interface. *J. Chem. Theory Comput.* **21**, 3649–3657 (2025).

71. Langer, M. F., Knoop, F., Carbogno, C., Scheffler, M. & Rupp, M. Heat flux for semilocal machine-learning potentials. *Phys. Rev. B* **108**, L100302 (2023).

72. Boone, P., Babaei, H. & Wilmer, C. E. Heat Flux for Many-Body Interactions: Corrections to LAMMPS. *J. Chem. Theory Comput.* **15**, 5579–5587 (2019).

73. Kubo, R. The fluctuation-dissipation theorem. *Rep. Prog. Phys.* **29**, 255 (1966).

74. Hardy, R. J. Energy-Flux Operator for a Lattice. *Phys. Rev.* **132**, 168–177 (1963).

75. Mandal, S. & Maiti, P. K. Prediction of thermal conductivity in CALF-20 with first-principles accuracy via machine learning interatomic potentials. *Commun. Mater.* **6**, 22 (2025).

76. Howell, P. C. Comparison of molecular dynamics methods and interatomic potentials for calculating the thermal conductivity of silicon. *J. Chem. Phys.* **137**, 224111 (2012).

77. Chen, J., Zhang, G. & Li, B. How to improve the accuracy of equilibrium molecular dynamics for computation of thermal conductivity? *Phys. Lett. A* **374**, 2392–2396 (2010).

78. Li, B., Wen, H.-M., Zhou, W. & Chen, B. Porous Metal–Organic Frameworks for Gas Storage and Separation: What, How, and Why? *J. Phys. Chem. Lett.* **5**, 3468–3479 (2014).

79. Babaei, H. & Wilmer, C. E. Mechanisms of Heat Transfer in Porous Crystals Containing Adsorbed Gases: Applications to Metal-Organic Frameworks. *Phys. Rev. Lett.* **116**, 025902 (2016).

80. Fan, H., Li, Z. & Zhou, Y. Influence of adsorbed gas molecules on thermal transport in metal-organic framework HKUST-1. *Phys. Rev. Appl.* **23**, 054006 (2025).

81. Thakur, S. & Giri, A. Impact of carbon dioxide loading on the thermal conductivity of metal organic frameworks. *J. Chem. Phys.* **162**, 154501 (2025).

82. Wiener, N. Generalized harmonic analysis. *Acta Math.* **55**, 117–258 (1930).

83. Khintchine, A. Korrelationstheorie der stationären stochastischen Prozesse. *Math. Ann.* **109**, 604–615 (1934).

84. Burnham, K. P. & Anderson, D. R. Multimodel Inference: Understanding AIC and BIC in Model Selection. *Sociol. Methods Res.* **33**, 261–304 (2004).

85. Cui, B. *et al.* Thermal Conductivity of ZIF-8 Thin-Film under Ambient Gas Pressure. *ACS Appl. Mater. Interfaces* **9**, 28139–28143 (2017).